\documentclass[aps,prl,reprint,superscriptaddress,paper=portrait]{revtex4-1}

\usepackage{graphicx}
\usepackage{amsmath}
\usepackage{amssymb}
\usepackage{xcolor}

\usepackage{siunitx}
\usepackage{ mathrsfs }
\usepackage{pdfpages}

\makeatletter
\AtBeginDocument{\let\LS@rot\@undefined}
\makeatother

\newcommand{\indts}{\ensuremath{{\bf j}}}
\newcommand{\indbes}{\ensuremath{m}}

\newcommand{\nbar}{\ensuremath{\bar{\mathrm{n}}}}
\newcommand{\nbarI}{\ensuremath{\nbar_\indts}}
\newcommand{\nbaro}{\ensuremath{\nbar_\text{0}}}
\newcommand{\nbari}{\ensuremath{\nbar_\text{1}}}
\newcommand{\nbarii}{\ensuremath{\nbar_\text{2}}}

\newcommand{\modind}{\ensuremath{\eta}}
\newcommand{\modindI}{\ensuremath{\modind_\indts}}
\newcommand{\modindo}{\ensuremath{\modind_\text{0}}}
\newcommand{\modindi}{\ensuremath{\modind_\text{1}}}

\newcommand{\exc}{\ensuremath{\mathcal{E}}}
\newcommand{\mo}{\ensuremath{\mathcal{M}}}

\newcommand{\dphmod}{\ensuremath{\Delta\varphi^\text{\mo}}}
\newcommand{\dphmodi}{\ensuremath{\dphmod_\text{1}}}
\newcommand{\dphmodoi}{\ensuremath{\Delta\varphi^\text{\mo}_\text{01}}}
\newcommand{\dphexc}{\ensuremath{\Delta\varphi^{\text{\exc}}_\text{02}}}
\newcommand{\oex}{\ensuremath{\Omega_\text{AC}}}
\newcommand{\oextheo}{\ensuremath{\Omega_\text{C}}}
\newcommand{\oexeff}{\ensuremath{\Omega_\text{C,eff}}}

\newcommand{\CPosc}{\ensuremath{\Pot^\text{osc}}}
\newcommand{\CPmod}{\ensuremath{\Pot^\text{\mo}}}
\newcommand{\CPexc}{\ensuremath{\Pot^\exc}}
\newcommand{\Pot}{\ensuremath{\phi}}
\newcommand{\phexc}{\ensuremath{\varphi^\exc}}
\newcommand{\phexco}{\ensuremath{\phexc_0}}
\newcommand{\phexcii}{\ensuremath{\phexc_2}}

\newcommand{\phmodi}{\ensuremath{\varphi^\mo_\text{1}}}

\newcommand{\phosc}{\ensuremath{\varphi^\text{osc}}}

\newcommand{\tramp}{\ensuremath{t_\text{ramp}}}
\newcommand{\trampphase}{\ensuremath{t_\text{$\varphi$,ramp}}}
\newcommand{\tcpl}{\ensuremath{t^\text{\mo}}}
\newcommand{\tcpli}{\ensuremath{t^\text{\mo}_\text{1}}}

\newcommand{\tcploi}{\ensuremath{t^\text{\mo}_\text{01}}}

\newcommand{\texco}{\ensuremath{t^\text{\exc}_\text{0}}}

\newcommand{\texci}{\ensuremath{t^\text{\exc}_\text{1}}}
\newcommand{\tpc}{\ensuremath{t^\text{\mo}_\text{1,prep}}}
\newcommand{\tpi}{\ensuremath{t_\pi}}

\newcommand{\TI}{\ensuremath{\text{T}_\indts}}
\newcommand{\To}{\ensuremath{\text{T}_\text{0}}}
\newcommand{\Ti}{\ensuremath{\text{T}_\text{1}}}
\newcommand{\Tii}{\ensuremath{\text{T}_\text{2}}}

\newcommand{\ucplI}{\ensuremath{u^{\text{\mo}}_\indts}}
\newcommand{\ucpli}{\ensuremath{u^{\text{\mo}}_\text{1}}}

\newcommand{\uexci}{\ensuremath{u^{\text{\exc}}_\text{1}}}
\newcommand{\uosc}{\ensuremath{u^\text{osc}}}

\newcommand{\wcpl}{\ensuremath{\Omega^\text{\mo}}}

\newcommand{\wcplpii}{\ensuremath{\wcpl_\text{1}/(2\pi)}}

\newcommand{\wcplpiI}{\ensuremath{\wcpl_\indts/(2\pi)}}
\newcommand{\wcplo}{\ensuremath{\wcpl_\text{0}}}
\newcommand{\wcpli}{\ensuremath{\wcpl_\text{1}}}
\newcommand{\wcplI}{\ensuremath{\wcpl_\indts}}

\newcommand{\wexc}{\ensuremath{\Omega^\text{\exc}}}
\newcommand{\wexci}{\ensuremath{\wexc_\text{1}}}

\newcommand{\wosc}{\ensuremath{\Omega^\text{osc}}}

\newcommand{\wlf}{\ensuremath{\omega}}

\newcommand{\wlfI}{\ensuremath{\omega_\indts}}
\newcommand{\wlfo}{\ensuremath{\wlf_\text{0}}}
\newcommand{\wlfi}{\ensuremath{\wlf_\text{1}}}

\newcommand{\wlfpii}{\ensuremath{\wlf_{\text{1}}/(2\pi)}}

\begin{document}
	
	
\title{Floquet-engineered vibrational dynamics in a two-dimensional array of trapped ions}
	
	
\author{Philip~Kiefer}
\email{philip.kiefer@physik.uni-freiburg.de}
\homepage{https://www.qsim.uni-freiburg.de}
\affiliation{Albert-Ludwigs-Universit\"at Freiburg, Physikalisches Institut, Hermann-Herder-Strasse 3, 79104 Freiburg, Germany}
\author{Frederick~Hakelberg}
\affiliation{Albert-Ludwigs-Universit\"at Freiburg, Physikalisches Institut, Hermann-Herder-Strasse 3, 79104 Freiburg, Germany}
\author{Matthias~Wittemer}
\affiliation{Albert-Ludwigs-Universit\"at Freiburg, Physikalisches Institut, Hermann-Herder-Strasse 3, 79104 Freiburg, Germany}
\author{Alejandro~Berm\'udez}
\email{albermud@fucm.es}
\affiliation{Departamento de F\'isica Te\'orica, Universidad Complutense, 28040 Madrid, Spain}
\author{Diego~Porras}
\email{d.porras@iff.csic.es}
\affiliation{Instituto de F\'isica Fundamental IFF-CSIC, Calle Serrano 113b, 28006 Madrid, Spain}
\author{Ulrich~Warring}
\affiliation{Albert-Ludwigs-Universit\"at Freiburg, Physikalisches Institut, Hermann-Herder-Strasse 3, 79104 Freiburg, Germany}
\author{Tobias~Schaetz}
\affiliation{Albert-Ludwigs-Universit\"at Freiburg, Physikalisches Institut, Hermann-Herder-Strasse 3, 79104 Freiburg, Germany}

	
	\date{\today}
	
	\begin{abstract}
	We demonstrate Floquet engineering in a basic yet scalable 2D architecture of individually trapped and controlled ions. 
	Local parametric modulations of detuned trapping potentials steer the strength of long-range inter-ion couplings and the related Peierls phase of the motional state. 
	In our proof-of-principle, we initialize large coherent states and tune modulation parameters to control trajectories, directions and interferences of the phonon flow.
	Our findings open a new pathway for future Floquet-based trapped-ion quantum simulators targeting correlated topological phenomena and dynamical gauge fields.
	\end{abstract}
	
	
	\maketitle


A promising route for the exploration of complex quantum dynamics is to use experimental simulator devices where synthetic interactions and quantum states can be efficiently controlled\,\cite{cirac_goals_2012}.
In general, systems of interest should provide long-range interactions and spatial dimensions higher than one since these remain beyond the reach of numerical methods \,\cite{verstraete_matrix_2008}.
A variety of prototype platforms already exists\,\cite{georgescu_quantum_2014}.
Trapped atomic ions are a promising approach, featuring identical constituents, long range Coulomb forces, and unique control of internal (electronic) and external (phonon) degrees of freedom\,\cite{ballance_high-fidelity_2016, gaebler_high-fidelity_2016}. 
Tremendous progress in common trapping potentials has been achieved\,\cite{zhang_observation_2017, jordan_near_2019}.
Furthermore experiments have shown coupling of individual ions at distant sites by matching local motional frequencies in 1D\,\cite{brown_coupled_2011,harlander_trapped-ion_2011,wilson_tunable_2014}  and in scalable 2D arrangements\,\cite{hakelberg_interference_2018} with the perspective to preserve the unique control of one/few ion ensembles. 
Typically, quantized vibrations (phonons) are used as an auxiliary bus mediating entangling gate operations\,\cite{schafer_fast_2018}, or synthetic spin-spin interactions\,\cite{schmitz_arch_2009}. 
In contrast, it has been proposed to actively use this degree of freedom. 
For example, to simulate complex bosonic lattice models\,\cite{porras_effective_2004} and to Floquet engineer an effective Peierls phase of the motional state, analogous to a synthetic gauge field\,\cite{bermudez_synthetic_2011-1, bermudez_photon-assisted-tunneling_2012}. 
In this context, phonons, represent charged particles in external electromagnetic fields.
They tunnel, their trajectories enclose areas related to geometric phases or interfere directly between individually controlled ions, located at distinct sites of a dedicated lattice structure. 
A realization requires fine tuning and parametric modulations of motional frequencies at each site. 
The strength of these drives tune the tunneling (coupling) strength, while control of the relative phases controls the accumulated Peierls phase. 
Such modulations enable inter-ion couplings between detuned (decoupled) trapping potentials by absorption and emission of energy (photons or phonons) out of the classical driving field.
Certain aspects of Floquet engineering by periodic modulations\,\cite{eckardt_colloquium_2017} have already been demonstrated for ultra-cold atoms\,\cite{aidelsburger_experimental_2011, struck_tunable_2012, asteria_measuring_2019}, superconducting qubits\,\cite{roushan_chiral_2017}, and photonic lattices\,\cite{mukherjee_experimental_2018}. 

In this Letter, we show essential features of phonon assisted coupling of individual atomic ions trapped at micro sites of our triangular two-dimensional trap array. 
Parametric drives applied to single or multiple locations steer constructive and destructive coherent couplings within the array.
Tuning driving amplitudes and relative phases, we control directionality and interference of the phonon flow via the related synthetic Peierls phase of the motional states, a key requirement for future quantum Floquet engineering.

We trap magnesium ions in our surface-electrode trap array featuring separate micro sites \TI, where  
$\indts \in\left\{0,1,2\right\}$  labels the  corners of the  triangle with side lengths of \SI{40}{\micro\meter} and an  ion-surface distance of $\simeq\SI{40}{\micro\meter}$\,\cite{mielenz_arrays_2016, sup}.
In harmonic approximation, these distances yield an inter-site coupling strength  $\oextheo/(2\pi)\simeq\SI{1}{\kilo\hertz}$ for motional frequencies $\omega_{\indts}/(2\pi) \simeq \SI{4}{\mega\hertz}$.
Heating rates of \SIrange{1}{10}{quanta/\milli\second} are derived from calibration measurements  close to the motional ground state\,\cite{ kalis_initialization_2017, friedenauer_high_2006}. 
In order to receive unambiguous signals of the phonon dynamics we initialize coherent states (exceeding $\simeq\SI{1000}{\ motional\ quanta}$).
The effective coupling rate \oexeff\ is tunable via relative motional mode orientations and the detuning of the individual trapping sites. 
Anharmonic contributions  of the trapping potential lead to an increased  $\oexeff/(2\pi)\simeq\SI{6}{\kilo\hertz}$,  while the related  efficiency is reduced accordingly\,\cite{hakelberg_interference_2018}.
Quasi-static control potentials  locally tune electric fields, curvatures and higher order  terms, e.g. for preparation/detection or inter-site couplings\,\cite{hakelberg_interference_2018}.
We adiabatically ramp these control potentials within $\tramp\leq\SI{100}{\micro\second}$ ($\tramp\geq(\omega_\indts)^{-1}$) between different configurations.
Additionally, we can apply various local periodic control potentials ${\CPosc_\indts} ({\wosc_\indts},{\phosc_\indts},{\uosc_\indts})$ for duration $t^{\text{osc}}_\indts$, oscillating with frequency $\Omega^{\rm osc}_\indts$ with a tunable phase $\phosc_\indts$\ and amplitude $\uosc_\indts$, allowing for: 
{\it(i)} Floquet engineering by a parametric modulation ${\CPmod_\indts}$ of motional frequencies with $\wcpl_\indts/(2\pi)\simeq\SI{100}{\kilo\hertz}$ or
{\it (ii)} Initialization of a vibronic coherent state at a given \TI\ by a local excitation via $\CPexc_\indts$.
All experiments are initialized by global Doppler cooling, aligning the local mode orientations, and tuning the lowest motional frequencies to $\omega_{\indts}/(2\pi) \simeq \SIrange{3}{5}{\mega\hertz}$\,\cite{mielenz_arrays_2016,hakelberg_interference_2018}. 
In following steps (see below) we apply dedicated control potentials for individual settings, and finish by local fluorescence detection allowing to derive average phonon numbers \nbarI. 
The reduction of the fluorescence rate  allows us to derive the increase of $\nbarI$. 
Each sequence is repeated 200 to 400 times to derive the standard error of the mean (s.e.m.).

\begin{figure}[h]
	\includegraphics{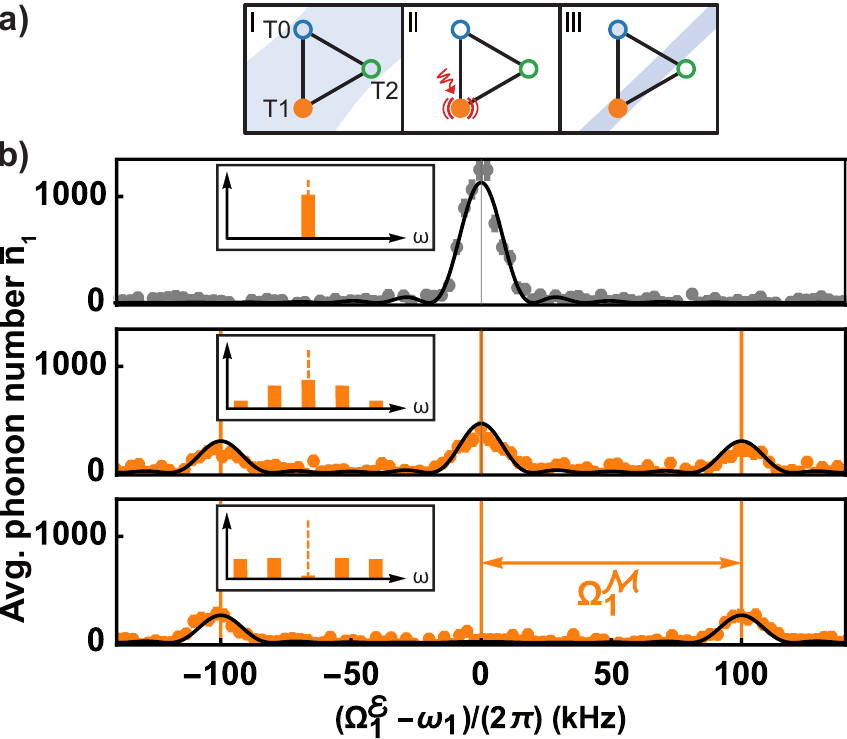}
	\caption{
		{\bf Characterization of the ion's frequency spectrum during modulation at a single site.}
		a) Experimental sequence to characterize the effect of the modulation at \Ti\ on the motional eigenfrequency \wlfi\ and its sidebands.
		These sidebands are interpreted as channels that in case of spectral overlap can permit transmission between neighboring \TI.
		The three, equidistant trapping sites are marked by colored circles, the presence of an ion by a disk.			
		(I) A single ion at \Ti\ is prepared via laser cooling (shaded area). 
		(II) The effect of the modulation potential $\CPmod_1$ with variable strength (red brackets) is probed simultaneously via the excitation field $\CPexc_1(\wexci)$ (red wiggled arrow)  for duration $\texci=\tcpli=\SI{50}{\micro\second}$. 
		(III) Local detection of fluorescence from \Ti\, enables reconstruction of the ion's motional amplitude. 
		(b) Final motional amplitudes (data points, errorbars s.e.m.) in dependence on excitation frequency \wexci: 
		(top) no modulation and 
		(middle, bottom) increasing modulation amplitude. 
		Model fits to the data (solid lines, see\,\cite{sup}) allow to calibrate the reference coherent excitation, i.e. $\modindi=0$  (top) and modulation indices with statistic uncertainties $\modindi=1.26(3)$, 2.57(4) (middle, bottom).
		Bar charts (insets) illustrate the derived amplitudes for the equally spaced spectral channels, shown up to the second sidebands.
		In this way, we can, e.g., strongly suppress excitations at the local oscillator frequency \wlfi (b, bottom).
		\label{fig:i}
	} 
\end{figure}

To calibrate Floquet engineering via oscillating potentials $\CPmod_\indts$, see\,\cite{sup}, we perform measurements with single ions at  \TI.
Exemplarily, we discuss results for site \Ti, see Fig.\,1, where we probe the effect of $\CPmod_{1}(\wcpli ,\ucpli)$ for fixed $\wcplpii=\SI{100}{kHz}\ll\wlfpii$ with a simultaneously applied drive for coherent excitation $\CPexc_{1}(\wexci ,\uexci)$.
Tuning \wexci\, across ${\wlfi}$, we show reconstructed motional amplitudes for $\ucpli =\SI{0}{\volt}$ (top), \SI{150}{\milli\volt} (middle), and \SI{250}{\milli\volt} (bottom) in dependence on \wexci.
When $\CPmod_{1}$ is switched on, a comb structure is spanned by several channels (sidebands) at $\Omega^{\exc}_1 \simeq \omega_1+ \indbes \Omega^{\mo}_1$ with $\indbes\in\mathbb{Z}$.
Floquet theory (solid lines) predicts that the relative strength of these channels is defined by the $\indbes$-th order Bessel function of the first kind $\mathcal{J}_\indbes(\modindI)$, where $\modindI \propto \ucplI /\wcplI $ represents the modulation index\,\cite{sup}.
For increasing \ucpli\,(Fig.\,1(b) middle, bottom), the carrier channel decreases, until  $\mathcal{J}_0(\modindI)$ crosses zero. 
This channel gets effectively shut and can lead in following experiments where we couple neighboring sites to so-called coherent destruction of tunneling\,\cite{grifoni_driven_1998}.  
Overall, we find  $\modindI\wcplpiI$ of up to \SI{300}{\kilo\hertz} for $\wcplpiI\simeq\SIrange{50}{200}{\kilo\hertz}$.

\begin{figure*}[ht]
	\includegraphics{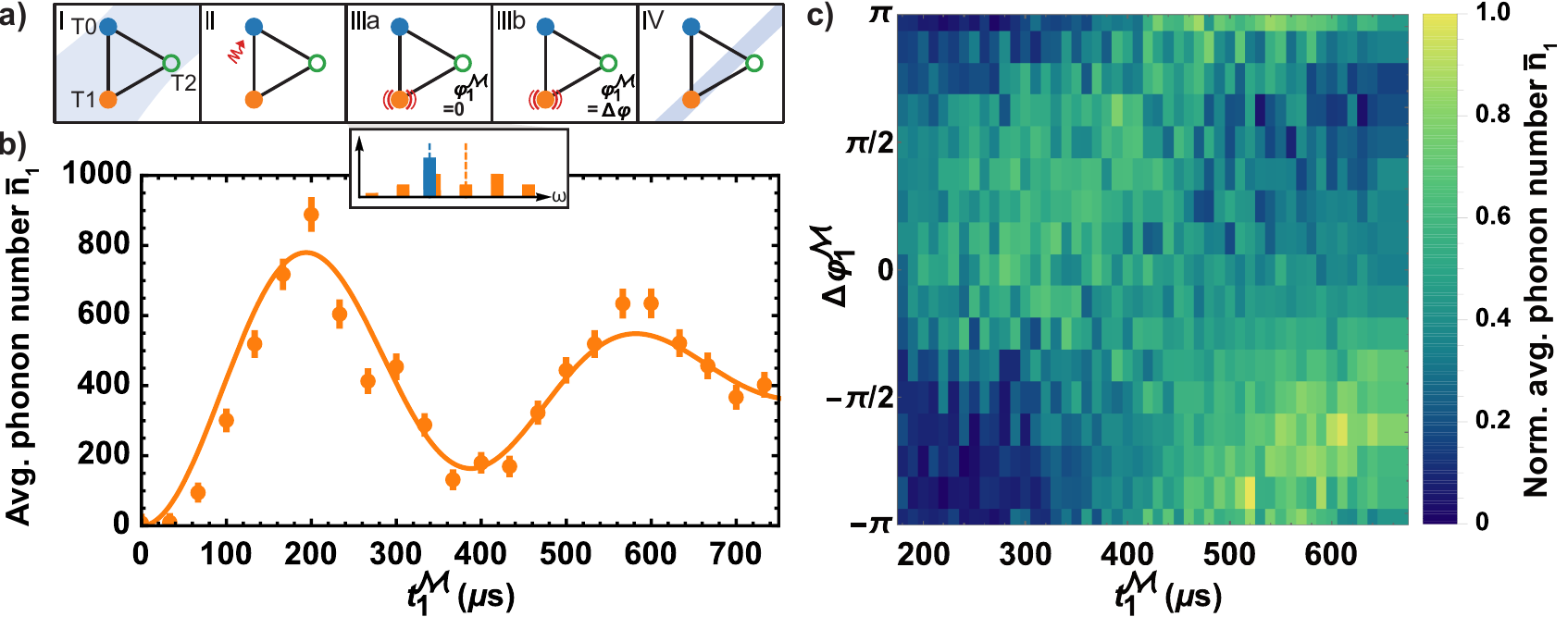}\\
	\caption{
		{\bf Inter-site coupling assisted by single-site modulation, enabling the  control of directionality of the phonon flow.}
		(a) Experimental sequence:
		(I) Initializing single ions at $\text{T}_0$ and $\text{T}_1$.
		(II) Switching to detuned coupling configuration and excite the ion at $\text{T}_0$.
		(IIIa) Launching inter-site coupling by application of $\CPmod_1$.
		(IIIb) To optionally redirect the phonon flow after \SI{150}{\micro\second}, we change the phase of $\CPmod_1$.
		Here $\dphmodi$ is realized within \SI{25}{\micro\second}, while $\CPmod_1$ is active in total for $\tcpli$.
		(IV) Detecting the motional amplitude \nbari.
		(b) Motional amplitude at \Ti\ as a function of \tcpli\ (data points, errorbars s.e.m.) using part (IIIa). 
		We choose $\CPmod_1$ to open efficiently the transmission channel of the first lower sideband at \Ti\ (inset, orange bar chart) and the carrier at \To\ (blue bar, slightly shifted for visibility).
		Phonons are coherently exchanged between $\text{T}_0$ and $\text{T}_1$.
		A model fit (solid line) yields an assisted coupling rate $\oex/(2\pi)=\SI{2.51(5)}{\kilo\hertz}$ and dephasing duration of $\tau = \SI{550(60)}{\micro\second}$.
		(c) Normalized motional amplitude in dependence on \tcpli\ and \dphmodi\ including sequence part (IIIb). 
		Real-time control of the Peierls phase $\Phi^{\rm P}(t)$ and the related directionality of the phonon flow is set by \dphmodi.
		\label{fig:ii}
}	\end{figure*}

To demonstrate control of synthetic Peierls phases imprinted on the motional state in real-time, we explore the assisted transfer of energy, i.e. flow of phonons, between ions at different \TI.
As an example we perform our experiment with single ions at \To\ and \Ti, see Fig.\,2.
We adjust the inter-site detuning to $\Delta\omega_{01}/(2\pi)=(\wlfi-\wlfo)/(2\pi)\simeq\SI{100}{\kilo\hertz}$, and excited the ion at \To\   ($\CPexc_0$ for $\texco=\SI{20}{\micro\second}$) to  $\nbaro=\SIrange{5000}{10000}{}$. 
By choice of $\Delta\omega_{01}$ the coupling efficiency (for $\modindI=0$) is suppressed by four orders of magnitude.
Setting $\wcpli=\Delta\omega_{01}$, $\modindi\simeq1.8$, at a fixed \phmodi, the phonon exchange is enabled by assistance of $\CPmod_1$:  
a single transmission channel is opened by overlapping the lower first sideband at \Ti\ with the carrier at \wlfo\ of \To, see Fig.\,2(b, inset).
In Figure\,2(b), we show reconstructed \nbari\ as a function of \tcpli.  
We model the coherent exchange, absorption and emission, of phonons, with corresponding assisted coupling rate $\oex$ (solid line, see\,\cite{sup}).
We thus confirm that the ion at $\text{T}_1$ absorbs up to $\simeq820$ phonons after $\tpi=\pi/\oex$.
The efficiency is limited to about \SI{15}{\percent} of \nbaro. We attribute that to the anharmonicity  of the trapping potential probed by the currently large \nbari\,\cite{hakelberg_interference_2018}. 
The anharmonic effects can be interpreted as additional detuning, increasing \oex\ but limiting efficiency.

The Peierls phase $\Phi^{\rm P}(t)$  plays an important role in the Floquet engineered Hamiltonian of energy transfer between different sites\,\cite{sup}. 
It is given by the path integral along $l$ between trapping sites
$\Phi^{\rm P}(t)=\frac{q}{\hbar}\int_{\TI}^{T_{\bf i}}{\rm d}\boldsymbol{l}\cdot \boldsymbol{A}(\boldsymbol{r},t)$, where $\hbar$ is the reduced Planck constant.
$\Phi^{\rm P}$ rules the dynamics of the phonons  as if they were  particles with charge $q$ coupled to a gauge potential $\boldsymbol{A}(\boldsymbol{r},t)$.
In an extended experimental sequence the evolution of $\Phi^{\rm P}$ is controlled in real-time during the experiment. 
In particular,  after a period of assisted coupling of $\tpc=\tpi/2\simeq\SI{150}{\micro\second}$, we adiabatically ramp  $\phmodi\to \phmodi+\dphmodi$ within $\trampphase=\SI{25}{\micro\second}\gg\wlfI^{-1}$, and continue the modulation for duration \tcpli-(\tpc+\trampphase). 
We depict normalized \nbari\ as a function of \dphmodi\ and \tcpli\ in Fig.\,2(c). 
The dependence of the  number of exchanged phonons on \dphmodi\  shows that  $\Phi^{\rm P}$  cannot be simply gauged away for $\dphmodi\neq0$ but, instead, it serves to control the directionality of the coherent energy flow. 
At  $\tcpli\simeq\SI{300}{\micro\second}$, optimal exchange is observed for $\dphmodi\simeq +\pi/4$. 
For $\dphmodi\approx-3\pi/4$, a change by $\pi$, we observe the transferred population nearly vanishing. 
This is consistent with Floquet theory, which predicts that $\oex(\phmodi)\to-\oex(\phmodi+\pi)$, equivalent to a time-reversal operation. That is, it returns phonons from \Ti\  back to  \To\ and further simulates the application of an electric field on charged particles.
In particular, $\partial\Phi^{\rm P}(t)/\partial t\neq 0$  can be understood as a background synthetic electric field $\boldsymbol{E}(\boldsymbol{r},t)=-\partial\boldsymbol{A}(\boldsymbol{r},t)/\partial t$.
We observe a global shift by $\pi/4$ in the data.
Numerical simulations can provide evidence for a similar shift, when considering a mismatch between \wcpli\ and $\Delta\omega_{01}$ of a few percent and the finite ramping duration \trampphase.	
\begin{figure}[h]
	\includegraphics{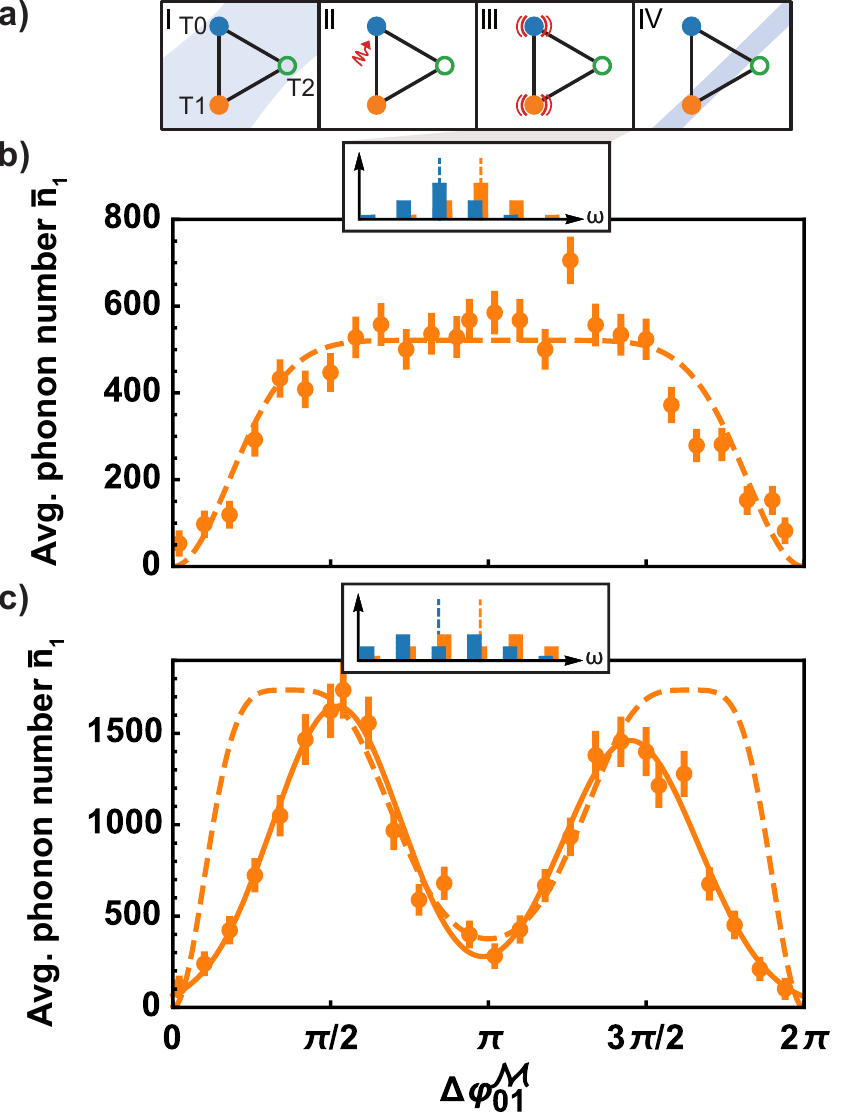}
	\caption{
		{\bf Tunable assisted coupling, established by selection of transmission channels and its dependence on the relative phase of two modulations.}
		(a) Experimental sequence:
		(I) Preparation of single ions at \To\ and  \Ti.
		(II) Initialization of phonons at \To\ by application of $\CPexc_0$ at detuned inter-site coupling configuration.
		(III) Activation of assisted coupling by $\CPmod_0$ and $\CPmod_1(\dphmodoi)$ with $\modind\simeq\modindo\simeq\modindi$ for $\tcpl =\tpi$, i.e. optimized exchange.
		(IV) Detection at \Ti.
		(b, c) Motional amplitude (data points, errorbars s.e.m.) in dependence on \dphmodoi\ for (b) $\modind\simeq0.9$ and (c) $\modind\simeq1.7$. 
		(Insets) Illustration of multiple transmission channels that mutually interfere and provide the effective coupling.
		The idealized, scaled model (dashed lines) describes the dependency qualitatively well, while it lacks to explain systematics in (c) outside the central region (guide to the eye: solid line), see text.
		Fundamentally, the energy transfer is a consequence of interferences of all  transmission channels and is tuned by \dphmodoi\, at $\modindo\simeq\modindi$.  
		\label{fig:iii}
	}
\end{figure}

In the next sequence, we explore the dynamics between  \To\ and \Ti\ when both sites are locally driven by  $\CPmod_0$ and $\CPmod_1$ at fixed $\wcplo=\wcpli\simeq\Delta\omega_{01}$ and $\modindo \simeq \modindi$. 
In this case, the assisted exchange  can be controlled by the relative modulation phase $\dphmodoi=\varphi_0^{\mathcal{M}}-\phmodi$, here reaching $\oex/(2\pi) \leq \SI{4.5}{\kilo\hertz}$\,\cite{hakelberg_interference_2018}.
As shown in  Fig.\,3,(b,c inset), we open several transmission channels, and the overall phonon exchange is governed by constructive or destructive interference of all contributions as a function of \dphmodoi.
We show results of \nbari\ after $\tcploi\simeq\tpi$ for $\modindo \simeq \modindi \simeq \modind \simeq \{0.9, 1.7\}$ in Figs.\,3(b) and (c), respectively.
For $\modind \simeq 0.9$, transmission is predominantly enabled by the resonance of two distinct channels, corresponding to  the carriers and first sidebands at \To\ and \Ti, cf. Fig.\,3(b, inset).  
As shown in Fig.\,3(b), the measured data is consistent with the  Floquet-engineered \oex, considering a linear coupling between harmonic oscillators (dashed line)\,\cite{bermudez_photon-assisted-tunneling_2012}: 
\oex\ and the transfer to \Ti\ is maximal and robust around $\dphmodoi=\pi$, while it is significantly suppressed for $\dphmodoi=0$ and 2$\pi$. 
A residual coupling for these values can be explained by a residual mismatch of the modulation indices ($\modindo \neq \modindi$) and inter-site dephasing.
Stronger modulation, see Fig.\,3(c,inset) opens additional channels, i.e., leads to larger contribution of upper and lower sidebands.
In Figure 3(c) data shows additional features of phase dependent energy transfer, e.g. an additional destructive interference near $\dphmodoi=\pi$ in accordance with the prediction. 
We note, however, that the two peaks of maximal phonon exchange  are narrowed with respect to the idealized theory, and a slight asymmetry appears, which we attribute to anharmonicities resulting from the Coulomb interaction, as well as local trapping potentials.
\begin{figure}[h!]
	\includegraphics{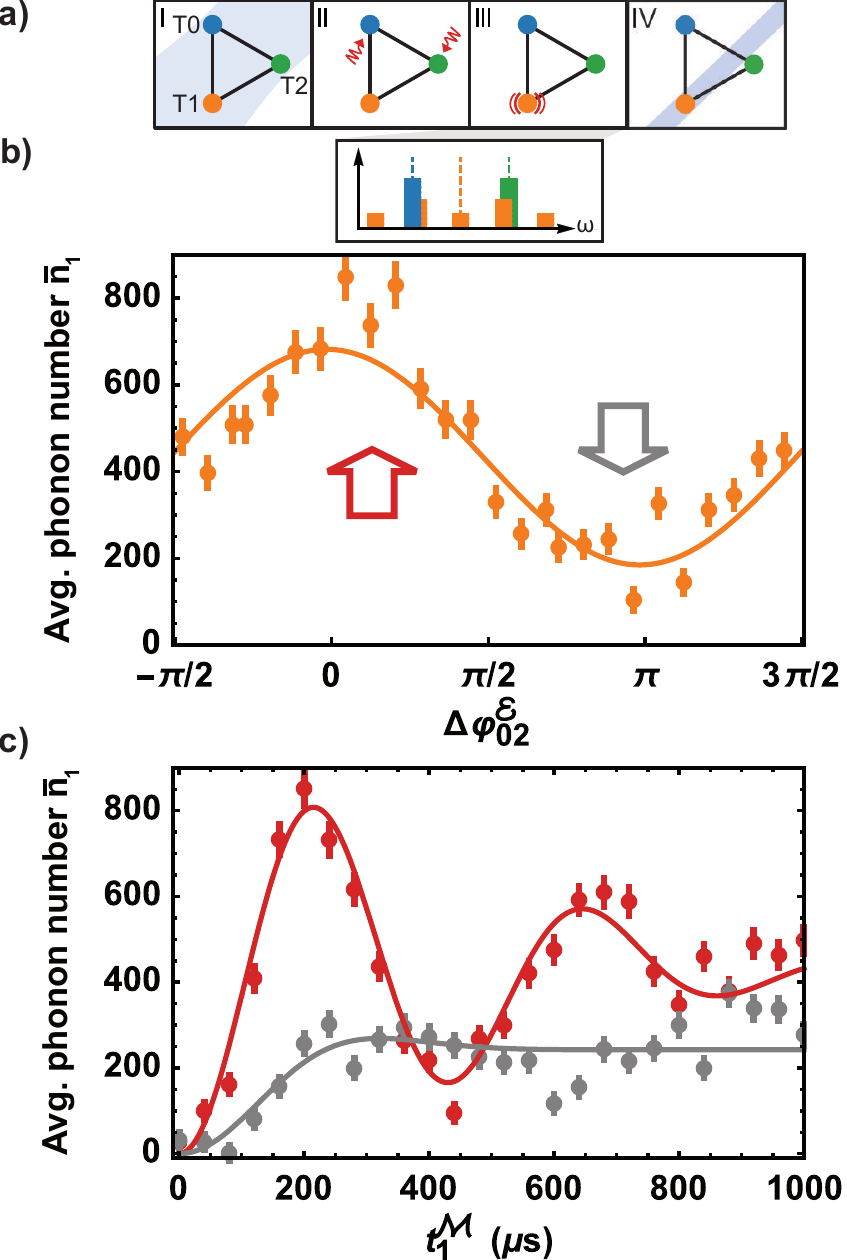}
	\caption{
		{\bf Controlled interference within a modulation-assisted 2D network of coupled oscillators.}
		(a) Experimental sequence:
		(I) Global preparation of single ions at \TI. 
		(II) Coupled network configuration offers a local gradient between the individual trapping sites $\Delta\omega_{01}=\Delta\omega_{12}$. Two local excitation pulses control the relative phase between coherent states at $\text{T}_0$ and $\text{T}_2$.  
		(III) Coupling the network via modulation $\CPmod_1$ for \tcpli.
		(IV) Detection at \Ti.
		(b, top) Modulation $\CPmod_1$ opens transmission channels between $\text{T}_0$ and $\text{T}_1$ as well as between $\text{T}_1$ and $\text{T}_2$.
		(b, bottom) Average phonon number \nbari\ as a function of \dphexc\ for $\tcpl=\SI{200}{\micro\second}$ (data points, errorbars s.e.m.). The  orange line represents a sinusoidal model fit yielding an amplitude of 250(20) quanta and an offset of 430(20) quanta.
		(c) Evolution for characteristic constructive (red) and destructive (gray) interference within the network.
		Model fits (solid lines) depict suppressed  \oex\ in the destructive case and dephasing durations of \SI{620(50)}{\micro\second} (red) and  \SI{250(80)}{\micro\second} (gray).
		Phase dependent dephasing durations as well as residual oscillations in the destructive case will be further investigated.
		\label{fig:iv}}
\end{figure}

To demonstrate  interference of phonons in two dimensions, akin to the Aharonov-Bohm effect of charged particles under an external magnetic field, all \TI\ are initialized.
We prepare for multi-site coupling by tuning  $\omega_{\left\{0,1,2\right\}}$/(2$\pi$) $\simeq$ \{4.9, 5.0, 5.1\} MHz.
We simultaneously apply $\CPexc_{0}(\phexco)$ and $\CPexc_{2}(\phexcii)$, to prepare coherent states with $\nbaro\simeq$ 5500  and $\nbarii\simeq$ 5800 phonons and, importantly, a  fixed phase relation \dphexc\,\footnote{Note: the phase relation of the ion oscillators results from different constant phase offsets, e.g. caused by supply wiring of the control electrodes, anharmonic contributions of the trapping potential or the duration between the start of excitation and modulation potentials.}. 
Multi-site phonon coupling is activated by applying $\CPmod_1$ with $\wcpli\simeq\Delta\wlf_{01}\simeq\Delta\wlf_{12}$ and $\modindi\simeq1.8$ during \tcpli. 
The lower and upper first sideband at $\text{T}_1$ opens transmission channels with the carrier at site \To\ and \Tii, respectively, see Fig.\,4(b, top).
We note that direct phonon exchange between  \To\ and \Tii\  is disabled by the frequency mismatch $\Delta\wlf_{02}\gg \oex$.
Results in Fig.\,4(b) depict  \nbari\  as a function of \dphexc\ for the maximal exchange achieved at  $\tcpli=\tpi$. 
While the energy transfer is maximal at $\dphexc=0$ (constructive interference highlighted by red arrow), it is minimal at $\dphexc=\pi$  (destructive interference highlighted by gray arrow). 
In Figure\,4(c), we investigate both of the extremal  settings in dependence on \tcpli, depicting the coherent destruction of energy transfer in 2D. 

To summarize, we demonstrate Floquet engineering of vibrational excitations in a 2D ion-trap array, present clear signatures of interference effects, and discuss the role of the arising dynamical Peierls phase.
In future studies, argon-ion bombardment\cite{hite_100-fold_2012} or cryogenic environments\,\cite{labaziewicz_suppression_2008} can reduce heating rates by more than two orders of magnitude permitting operation near the motional ground state for durations $\gg 1/\oex$, as established for short time scales already\,\cite{kalis_initialization_2017}. 
Furthermore triangular lattices, plaquettes to concatenate rhombic ladders and even more complex, non-periodic structures can be realized in future arrays\,\cite{schaetz_towards_2007, schaetz_focus_2013}.
Combining these techniques with the presented Floquet toolbox may additionally enable to study the interplay of non-linearities, i.e. effective on-site phonon-phonon interaction, with the synthetic gauge fields. 
This would enable  to explore correlated, symmetry-protected topological phases of bosons\,\cite{huber_topological_2011}.   
By exploiting laser cooling and heating mechanisms\,\cite{lemmer_trapped-ion_2018}, we can build a phononic analog of photonic lattices\,\cite{bermudez_controlling_2013,peano_topological_2016,ozawa_topological_2019,porras_topological_2019} with a rich interplay between topological and dissipative effects. 
Application of state-dependent optical potentials, may further extend the quantum-simulation prospects of our platform by enabling state dependent transmission. 
In particular, by coupling the internal degrees of freedom to the vibrations, one may study bosonic lattice models in the presence of dynamical gauge fields, e.g. famous Aharonov-Bohm physics, cages and edge states\,\cite{bermudez_photon-assisted-tunneling_2012, bermudez_interaction-dependent_2015}.

\begin{acknowledgments}
We thank J.-P. Schr\"oder for help with the experimental control system. The trap chip was designed in collaboration with R. Schmied in a cooperation with the NIST ion storage group and produced by Sandia National Laboratories. This work was supported by the Deutsche Forschungsgemeinschaft (DFG) [SCHA 973/6-3].\\
P.K. and F.H. contributed equally to this work.
\end{acknowledgments}
	

\begin{thebibliography}{39}%
	\makeatletter
	\providecommand \@ifxundefined [1]{%
		\@ifx{#1\undefined}
	}%
	\providecommand \@ifnum [1]{%
		\ifnum #1\expandafter \@firstoftwo
		\else \expandafter \@secondoftwo
		\fi
	}%
	\providecommand \@ifx [1]{%
		\ifx #1\expandafter \@firstoftwo
		\else \expandafter \@secondoftwo
		\fi
	}%
	\providecommand \natexlab [1]{#1}%
	\providecommand \enquote  [1]{``#1''}%
	\providecommand \bibnamefont  [1]{#1}%
	\providecommand \bibfnamefont [1]{#1}%
	\providecommand \citenamefont [1]{#1}%
	\providecommand \href@noop [0]{\@secondoftwo}%
	\providecommand \href [0]{\begingroup \@sanitize@url \@href}%
	\providecommand \@href[1]{\@@startlink{#1}\@@href}%
	\providecommand \@@href[1]{\endgroup#1\@@endlink}%
	\providecommand \@sanitize@url [0]{\catcode `\\12\catcode `\$12\catcode
		`\&12\catcode `\#12\catcode `\^12\catcode `\_12\catcode `\%12\relax}%
	\providecommand \@@startlink[1]{}%
	\providecommand \@@endlink[0]{}%
	\providecommand \url  [0]{\begingroup\@sanitize@url \@url }%
	\providecommand \@url [1]{\endgroup\@href {#1}{\urlprefix }}%
	\providecommand \urlprefix  [0]{URL }%
	\providecommand \Eprint [0]{\href }%
	\providecommand \doibase [0]{http://dx.doi.org/}%
	\providecommand \selectlanguage [0]{\@gobble}%
	\providecommand \bibinfo  [0]{\@secondoftwo}%
	\providecommand \bibfield  [0]{\@secondoftwo}%
	\providecommand \translation [1]{[#1]}%
	\providecommand \BibitemOpen [0]{}%
	\providecommand \bibitemStop [0]{}%
	\providecommand \bibitemNoStop [0]{.\EOS\space}%
	\providecommand \EOS [0]{\spacefactor3000\relax}%
	\providecommand \BibitemShut  [1]{\csname bibitem#1\endcsname}%
	\let\auto@bib@innerbib\@empty
	\bibitem [{\citenamefont {Cirac}\ and\ \citenamefont
		{Zoller}(2012)}]{cirac_goals_2012}%
	\BibitemOpen
	\bibfield  {author} {\bibinfo {author} {\bibfnamefont {J.~I.}\ \bibnamefont
			{Cirac}}\ and\ \bibinfo {author} {\bibfnamefont {P.}~\bibnamefont {Zoller}},\
	}\href {\doibase 10.1038/nphys2275} {\bibfield  {journal} {\bibinfo
			{journal} {Nat. Phys.}\ }\textbf {\bibinfo {volume} {8}},\ \bibinfo {pages}
		{264} (\bibinfo {year} {2012})}\BibitemShut {NoStop}%
	\bibitem [{\citenamefont {Verstraete}\ \emph {et~al.}(2008)\citenamefont
		{Verstraete}, \citenamefont {Cirac},\ and\ \citenamefont
		{Murg}}]{verstraete_matrix_2008}%
	\BibitemOpen
	\bibfield  {author} {\bibinfo {author} {\bibfnamefont {F.}~\bibnamefont
			{Verstraete}}, \bibinfo {author} {\bibfnamefont {J.~I.}\ \bibnamefont
			{Cirac}}, \ and\ \bibinfo {author} {\bibfnamefont {V.}~\bibnamefont {Murg}},\
	}\href {\doibase 10.1080/14789940801912366} {\bibfield  {journal} {\bibinfo
			{journal} {Advances in Physics}\ }\textbf {\bibinfo {volume} {57}},\ \bibinfo
		{pages} {143} (\bibinfo {year} {2008})}\BibitemShut {NoStop}%
	\bibitem [{\citenamefont {Georgescu}\ \emph {et~al.}(2014)\citenamefont
		{Georgescu}, \citenamefont {Ashhab},\ and\ \citenamefont
		{Nori}}]{georgescu_quantum_2014}%
	\BibitemOpen
	\bibfield  {author} {\bibinfo {author} {\bibfnamefont {I.~M.}\ \bibnamefont
			{Georgescu}}, \bibinfo {author} {\bibfnamefont {S.}~\bibnamefont {Ashhab}}, \
		and\ \bibinfo {author} {\bibfnamefont {F.}~\bibnamefont {Nori}},\ }\href
	{\doibase 10.1103/RevModPhys.86.153} {\bibfield  {journal} {\bibinfo
			{journal} {Reviews of Modern Physics}\ }\textbf {\bibinfo {volume} {86}},\
		\bibinfo {pages} {153} (\bibinfo {year} {2014})}\BibitemShut {NoStop}%
	\bibitem [{\citenamefont {Ballance}\ \emph {et~al.}(2016)\citenamefont
		{Ballance}, \citenamefont {Harty}, \citenamefont {Linke}, \citenamefont
		{Sepiol},\ and\ \citenamefont {Lucas}}]{ballance_high-fidelity_2016}%
	\BibitemOpen
	\bibfield  {author} {\bibinfo {author} {\bibfnamefont {C.~J.}\ \bibnamefont
			{Ballance}}, \bibinfo {author} {\bibfnamefont {T.~P.}\ \bibnamefont {Harty}},
		\bibinfo {author} {\bibfnamefont {N.~M.}\ \bibnamefont {Linke}}, \bibinfo
		{author} {\bibfnamefont {M.~A.}\ \bibnamefont {Sepiol}}, \ and\ \bibinfo
		{author} {\bibfnamefont {D.~M.}\ \bibnamefont {Lucas}},\ }\href {\doibase
		10.1103/PhysRevLett.117.060504} {\bibfield  {journal} {\bibinfo  {journal}
			{Physical Review Letters}\ }\textbf {\bibinfo {volume} {117}},\ \bibinfo
		{pages} {060504} (\bibinfo {year} {2016})}\BibitemShut {NoStop}%
	\bibitem [{\citenamefont {Gaebler}\ \emph {et~al.}(2016)\citenamefont
		{Gaebler}, \citenamefont {Tan}, \citenamefont {Lin}, \citenamefont {Wan},
		\citenamefont {Bowler}, \citenamefont {Keith}, \citenamefont {Glancy},
		\citenamefont {Coakley}, \citenamefont {Knill}, \citenamefont {Leibfried},\
		and\ \citenamefont {Wineland}}]{gaebler_high-fidelity_2016}%
	\BibitemOpen
	\bibfield  {author} {\bibinfo {author} {\bibfnamefont {J.~P.}\ \bibnamefont
			{Gaebler}}, \bibinfo {author} {\bibfnamefont {T.~R.}\ \bibnamefont {Tan}},
		\bibinfo {author} {\bibfnamefont {Y.}~\bibnamefont {Lin}}, \bibinfo {author}
		{\bibfnamefont {Y.}~\bibnamefont {Wan}}, \bibinfo {author} {\bibfnamefont
			{R.}~\bibnamefont {Bowler}}, \bibinfo {author} {\bibfnamefont {A.~C.}\
			\bibnamefont {Keith}}, \bibinfo {author} {\bibfnamefont {S.}~\bibnamefont
			{Glancy}}, \bibinfo {author} {\bibfnamefont {K.}~\bibnamefont {Coakley}},
		\bibinfo {author} {\bibfnamefont {E.}~\bibnamefont {Knill}}, \bibinfo
		{author} {\bibfnamefont {D.}~\bibnamefont {Leibfried}}, \ and\ \bibinfo
		{author} {\bibfnamefont {D.~J.}\ \bibnamefont {Wineland}},\ }\href {\doibase
		10.1103/PhysRevLett.117.060505} {\bibfield  {journal} {\bibinfo  {journal}
			{Physical Review Letters}\ }\textbf {\bibinfo {volume} {117}},\ \bibinfo
		{pages} {060505} (\bibinfo {year} {2016})}\BibitemShut {NoStop}%
	\bibitem [{\citenamefont {Zhang}\ \emph {et~al.}(2017)\citenamefont {Zhang},
		\citenamefont {Pagano}, \citenamefont {Hess}, \citenamefont {Kyprianidis},
		\citenamefont {Becker}, \citenamefont {Kaplan}, \citenamefont {Gorshkov},
		\citenamefont {Gong},\ and\ \citenamefont {Monroe}}]{zhang_observation_2017}%
	\BibitemOpen
	\bibfield  {author} {\bibinfo {author} {\bibfnamefont {J.}~\bibnamefont
			{Zhang}}, \bibinfo {author} {\bibfnamefont {G.}~\bibnamefont {Pagano}},
		\bibinfo {author} {\bibfnamefont {P.~W.}\ \bibnamefont {Hess}}, \bibinfo
		{author} {\bibfnamefont {A.}~\bibnamefont {Kyprianidis}}, \bibinfo {author}
		{\bibfnamefont {P.}~\bibnamefont {Becker}}, \bibinfo {author} {\bibfnamefont
			{H.}~\bibnamefont {Kaplan}}, \bibinfo {author} {\bibfnamefont {A.~V.}\
			\bibnamefont {Gorshkov}}, \bibinfo {author} {\bibfnamefont {Z.-X.}\
			\bibnamefont {Gong}}, \ and\ \bibinfo {author} {\bibfnamefont
			{C.}~\bibnamefont {Monroe}},\ }\href {\doibase 10.1038/nature24654}
	{\bibfield  {journal} {\bibinfo  {journal} {Nature}\ }\textbf {\bibinfo
			{volume} {551}},\ \bibinfo {pages} {601} (\bibinfo {year}
		{2017})}\BibitemShut {NoStop}%
	\bibitem [{\citenamefont {Jordan}\ \emph {et~al.}(2019)\citenamefont {Jordan},
		\citenamefont {Gilmore}, \citenamefont {Shankar}, \citenamefont
		{{Safavi-Naini}}, \citenamefont {Bohnet}, \citenamefont {Holland},\ and\
		\citenamefont {Bollinger}}]{jordan_near_2019}%
	\BibitemOpen
	\bibfield  {author} {\bibinfo {author} {\bibfnamefont {E.}~\bibnamefont
			{Jordan}}, \bibinfo {author} {\bibfnamefont {K.~A.}\ \bibnamefont {Gilmore}},
		\bibinfo {author} {\bibfnamefont {A.}~\bibnamefont {Shankar}}, \bibinfo
		{author} {\bibfnamefont {A.}~\bibnamefont {{Safavi-Naini}}}, \bibinfo
		{author} {\bibfnamefont {J.~G.}\ \bibnamefont {Bohnet}}, \bibinfo {author}
		{\bibfnamefont {M.~J.}\ \bibnamefont {Holland}}, \ and\ \bibinfo {author}
		{\bibfnamefont {J.~J.}\ \bibnamefont {Bollinger}},\ }\href {\doibase
		10.1103/PhysRevLett.122.053603} {\bibfield  {journal} {\bibinfo  {journal}
			{Physical Review Letters}\ }\textbf {\bibinfo {volume} {122}},\ \bibinfo
		{pages} {053603} (\bibinfo {year} {2019})}\BibitemShut {NoStop}%
	\bibitem [{\citenamefont {Brown}\ \emph {et~al.}(2011)\citenamefont {Brown},
		\citenamefont {Ospelkaus}, \citenamefont {Colombe}, \citenamefont {Wilson},
		\citenamefont {Leibfried},\ and\ \citenamefont
		{Wineland}}]{brown_coupled_2011}%
	\BibitemOpen
	\bibfield  {author} {\bibinfo {author} {\bibfnamefont {K.~R.}\ \bibnamefont
			{Brown}}, \bibinfo {author} {\bibfnamefont {C.}~\bibnamefont {Ospelkaus}},
		\bibinfo {author} {\bibfnamefont {Y.}~\bibnamefont {Colombe}}, \bibinfo
		{author} {\bibfnamefont {A.~C.}\ \bibnamefont {Wilson}}, \bibinfo {author}
		{\bibfnamefont {D.}~\bibnamefont {Leibfried}}, \ and\ \bibinfo {author}
		{\bibfnamefont {D.~J.}\ \bibnamefont {Wineland}},\ }\href {\doibase
		10.1038/nature09721} {\bibfield  {journal} {\bibinfo  {journal} {Nature}\
		}\textbf {\bibinfo {volume} {471}},\ \bibinfo {pages} {196} (\bibinfo {year}
		{2011})}\BibitemShut {NoStop}%
	\bibitem [{\citenamefont {Harlander}\ \emph {et~al.}(2011)\citenamefont
		{Harlander}, \citenamefont {Lechner}, \citenamefont {Brownnutt},
		\citenamefont {Blatt},\ and\ \citenamefont
		{H{\"a}nsel}}]{harlander_trapped-ion_2011}%
	\BibitemOpen
	\bibfield  {author} {\bibinfo {author} {\bibfnamefont {M.}~\bibnamefont
			{Harlander}}, \bibinfo {author} {\bibfnamefont {R.}~\bibnamefont {Lechner}},
		\bibinfo {author} {\bibfnamefont {M.}~\bibnamefont {Brownnutt}}, \bibinfo
		{author} {\bibfnamefont {R.}~\bibnamefont {Blatt}}, \ and\ \bibinfo {author}
		{\bibfnamefont {W.}~\bibnamefont {H{\"a}nsel}},\ }\href {\doibase
		10.1038/nature09800} {\bibfield  {journal} {\bibinfo  {journal} {Nature}\
		}\textbf {\bibinfo {volume} {471}},\ \bibinfo {pages} {200} (\bibinfo {year}
		{2011})}\BibitemShut {NoStop}%
	\bibitem [{\citenamefont {Wilson}\ \emph {et~al.}(2014)\citenamefont {Wilson},
		\citenamefont {Colombe}, \citenamefont {Brown}, \citenamefont {Knill},
		\citenamefont {Leibfried},\ and\ \citenamefont
		{Wineland}}]{wilson_tunable_2014}%
	\BibitemOpen
	\bibfield  {author} {\bibinfo {author} {\bibfnamefont {A.~C.}\ \bibnamefont
			{Wilson}}, \bibinfo {author} {\bibfnamefont {Y.}~\bibnamefont {Colombe}},
		\bibinfo {author} {\bibfnamefont {K.~R.}\ \bibnamefont {Brown}}, \bibinfo
		{author} {\bibfnamefont {E.}~\bibnamefont {Knill}}, \bibinfo {author}
		{\bibfnamefont {D.}~\bibnamefont {Leibfried}}, \ and\ \bibinfo {author}
		{\bibfnamefont {D.~J.}\ \bibnamefont {Wineland}},\ }\href {\doibase
		10.1038/nature13565} {\bibfield  {journal} {\bibinfo  {journal} {Nature}\
		}\textbf {\bibinfo {volume} {512}},\ \bibinfo {pages} {57} (\bibinfo {year}
		{2014})}\BibitemShut {NoStop}%
	\bibitem [{\citenamefont {Hakelberg}\ \emph {et~al.}(2018)\citenamefont
		{Hakelberg}, \citenamefont {Kiefer}, \citenamefont {Wittemer}, \citenamefont
		{Warring},\ and\ \citenamefont {Schaetz}}]{hakelberg_interference_2018}%
	\BibitemOpen
	\bibfield  {author} {\bibinfo {author} {\bibfnamefont {F.}~\bibnamefont
			{Hakelberg}}, \bibinfo {author} {\bibfnamefont {P.}~\bibnamefont {Kiefer}},
		\bibinfo {author} {\bibfnamefont {M.}~\bibnamefont {Wittemer}}, \bibinfo
		{author} {\bibfnamefont {U.}~\bibnamefont {Warring}}, \ and\ \bibinfo
		{author} {\bibfnamefont {T.}~\bibnamefont {Schaetz}},\ }\href@noop {}
	{\bibfield  {journal} {\bibinfo  {journal} {arXiv:1812.08552}\ } (\bibinfo
		{year} {2018})}.\ \Eprint {http://arxiv.org/abs/1812.08552}
 \BibitemShut%
	\bibitem [{\citenamefont {Sch{\"a}fer}\ \emph {et~al.}(2018)\citenamefont
		{Sch{\"a}fer}, \citenamefont {Ballance}, \citenamefont {Thirumalai},
		\citenamefont {Stephenson}, \citenamefont {Ballance}, \citenamefont
		{Steane},\ and\ \citenamefont {Lucas}}]{schafer_fast_2018}%
	\BibitemOpen
	\bibfield  {author} {\bibinfo {author} {\bibfnamefont {V.~M.}\ \bibnamefont
			{Sch{\"a}fer}}, \bibinfo {author} {\bibfnamefont {C.~J.}\ \bibnamefont
			{Ballance}}, \bibinfo {author} {\bibfnamefont {K.}~\bibnamefont
			{Thirumalai}}, \bibinfo {author} {\bibfnamefont {L.~J.}\ \bibnamefont
			{Stephenson}}, \bibinfo {author} {\bibfnamefont {T.~G.}\ \bibnamefont
			{Ballance}}, \bibinfo {author} {\bibfnamefont {A.~M.}\ \bibnamefont
			{Steane}}, \ and\ \bibinfo {author} {\bibfnamefont {D.~M.}\ \bibnamefont
			{Lucas}},\ }\href {\doibase 10.1038/nature25737} {\bibfield  {journal}
		{\bibinfo  {journal} {Nature}\ }\textbf {\bibinfo {volume} {555}},\ \bibinfo
		{pages} {75} (\bibinfo {year} {2018})}\BibitemShut {NoStop}%
	\bibitem [{\citenamefont {Schmitz}\ \emph {et~al.}(2009)\citenamefont
		{Schmitz}, \citenamefont {Friedenauer}, \citenamefont {Schneider},
		\citenamefont {Matjeschk}, \citenamefont {Enderlein}, \citenamefont {Huber},
		\citenamefont {Glueckert}, \citenamefont {Porras},\ and\ \citenamefont
		{Schaetz}}]{schmitz_arch_2009}%
	\BibitemOpen
	\bibfield  {author} {\bibinfo {author} {\bibfnamefont {H.}~\bibnamefont
			{Schmitz}}, \bibinfo {author} {\bibfnamefont {A.}~\bibnamefont
			{Friedenauer}}, \bibinfo {author} {\bibfnamefont {C.}~\bibnamefont
			{Schneider}}, \bibinfo {author} {\bibfnamefont {R.}~\bibnamefont
			{Matjeschk}}, \bibinfo {author} {\bibfnamefont {M.}~\bibnamefont
			{Enderlein}}, \bibinfo {author} {\bibfnamefont {T.}~\bibnamefont {Huber}},
		\bibinfo {author} {\bibfnamefont {J.}~\bibnamefont {Glueckert}}, \bibinfo
		{author} {\bibfnamefont {D.}~\bibnamefont {Porras}}, \ and\ \bibinfo {author}
		{\bibfnamefont {T.}~\bibnamefont {Schaetz}},\ }\href {\doibase
		10.1007/s00340-009-3455-6} {\bibfield  {journal} {\bibinfo  {journal}
			{Applied Physics B}\ }\textbf {\bibinfo {volume} {95}},\ \bibinfo {pages}
		{195} (\bibinfo {year} {2009})}\BibitemShut {NoStop}%
	\bibitem [{\citenamefont {Porras}\ and\ \citenamefont
		{Cirac}(2004)}]{porras_effective_2004}%
	\BibitemOpen
	\bibfield  {author} {\bibinfo {author} {\bibfnamefont {D.}~\bibnamefont
			{Porras}}\ and\ \bibinfo {author} {\bibfnamefont {J.~I.}\ \bibnamefont
			{Cirac}},\ }\href@noop {} {\bibfield  {journal} {\bibinfo  {journal} {Phys.
				Rev. Lett.}\ }\textbf {\bibinfo {volume} {92}},\ \bibinfo {pages} {207901}
		(\bibinfo {year} {2004})}\BibitemShut {NoStop}%
	\bibitem [{\citenamefont {Bermudez}\ \emph {et~al.}(2011)\citenamefont
		{Bermudez}, \citenamefont {Schaetz},\ and\ \citenamefont
		{Porras}}]{bermudez_synthetic_2011-1}%
	\BibitemOpen
	\bibfield  {author} {\bibinfo {author} {\bibfnamefont {A.}~\bibnamefont
			{Bermudez}}, \bibinfo {author} {\bibfnamefont {T.}~\bibnamefont {Schaetz}}, \
		and\ \bibinfo {author} {\bibfnamefont {D.}~\bibnamefont {Porras}},\ }\href
	{\doibase 10.1103/PhysRevLett.107.150501} {\bibfield  {journal} {\bibinfo
			{journal} {Physical Review Letters}\ }\textbf {\bibinfo {volume} {107}},\
		\bibinfo {pages} {150501} (\bibinfo {year} {2011})}\BibitemShut {NoStop}%
	\bibitem [{\citenamefont {Bermudez}\ \emph {et~al.}(2012)\citenamefont
		{Bermudez}, \citenamefont {Schaetz},\ and\ \citenamefont
		{Porras}}]{bermudez_photon-assisted-tunneling_2012}%
	\BibitemOpen
	\bibfield  {author} {\bibinfo {author} {\bibfnamefont {A.}~\bibnamefont
			{Bermudez}}, \bibinfo {author} {\bibfnamefont {T.}~\bibnamefont {Schaetz}}, \
		and\ \bibinfo {author} {\bibfnamefont {D.}~\bibnamefont {Porras}},\ }\href
	{\doibase 10.1088/1367-2630/14/5/053049} {\bibfield  {journal} {\bibinfo
			{journal} {New Journal of Physics}\ }\textbf {\bibinfo {volume} {14}},\
		\bibinfo {pages} {053049} (\bibinfo {year} {2012})}\BibitemShut {NoStop}%
	\bibitem [{\citenamefont {Eckardt}(2017)}]{eckardt_colloquium_2017}%
	\BibitemOpen
	\bibfield  {author} {\bibinfo {author} {\bibfnamefont {A.}~\bibnamefont
			{Eckardt}},\ }\href {\doibase 10.1103/RevModPhys.89.011004} {\bibfield
		{journal} {\bibinfo  {journal} {Reviews of Modern Physics}\ }\textbf
		{\bibinfo {volume} {89}},\ \bibinfo {pages} {011004} (\bibinfo {year}
		{2017})}\BibitemShut {NoStop}%
	\bibitem [{\citenamefont {Aidelsburger}\ \emph {et~al.}(2011)\citenamefont
		{Aidelsburger}, \citenamefont {Atala}, \citenamefont {Nascimb{\`e}ne},
		\citenamefont {Trotzky}, \citenamefont {Chen},\ and\ \citenamefont
		{Bloch}}]{aidelsburger_experimental_2011}%
	\BibitemOpen
	\bibfield  {author} {\bibinfo {author} {\bibfnamefont {M.}~\bibnamefont
			{Aidelsburger}}, \bibinfo {author} {\bibfnamefont {M.}~\bibnamefont {Atala}},
		\bibinfo {author} {\bibfnamefont {S.}~\bibnamefont {Nascimb{\`e}ne}},
		\bibinfo {author} {\bibfnamefont {S.}~\bibnamefont {Trotzky}}, \bibinfo
		{author} {\bibfnamefont {Y.-A.}\ \bibnamefont {Chen}}, \ and\ \bibinfo
		{author} {\bibfnamefont {I.}~\bibnamefont {Bloch}},\ }\href {\doibase
		10.1103/PhysRevLett.107.255301} {\bibfield  {journal} {\bibinfo  {journal}
			{Physical Review Letters}\ }\textbf {\bibinfo {volume} {107}},\ \bibinfo
		{pages} {255301} (\bibinfo {year} {2011})}\BibitemShut {NoStop}%
	\bibitem [{\citenamefont {Struck}\ \emph {et~al.}(2012)\citenamefont {Struck},
		\citenamefont {{\"O}lschl{\"a}ger}, \citenamefont {Weinberg}, \citenamefont
		{Hauke}, \citenamefont {Simonet}, \citenamefont {Eckardt}, \citenamefont
		{Lewenstein}, \citenamefont {Sengstock},\ and\ \citenamefont
		{Windpassinger}}]{struck_tunable_2012}%
	\BibitemOpen
	\bibfield  {author} {\bibinfo {author} {\bibfnamefont {J.}~\bibnamefont
			{Struck}}, \bibinfo {author} {\bibfnamefont {C.}~\bibnamefont
			{{\"O}lschl{\"a}ger}}, \bibinfo {author} {\bibfnamefont {M.}~\bibnamefont
			{Weinberg}}, \bibinfo {author} {\bibfnamefont {P.}~\bibnamefont {Hauke}},
		\bibinfo {author} {\bibfnamefont {J.}~\bibnamefont {Simonet}}, \bibinfo
		{author} {\bibfnamefont {A.}~\bibnamefont {Eckardt}}, \bibinfo {author}
		{\bibfnamefont {M.}~\bibnamefont {Lewenstein}}, \bibinfo {author}
		{\bibfnamefont {K.}~\bibnamefont {Sengstock}}, \ and\ \bibinfo {author}
		{\bibfnamefont {P.}~\bibnamefont {Windpassinger}},\ }\href {\doibase
		10.1103/PhysRevLett.108.225304} {\bibfield  {journal} {\bibinfo  {journal}
			{Physical Review Letters}\ }\textbf {\bibinfo {volume} {108}},\ \bibinfo
		{pages} {225304} (\bibinfo {year} {2012})}\BibitemShut {NoStop}%
	\bibitem [{\citenamefont {Asteria}\ \emph {et~al.}(2019)\citenamefont
		{Asteria}, \citenamefont {Tran}, \citenamefont {Ozawa}, \citenamefont
		{Tarnowski}, \citenamefont {Rem}, \citenamefont {Fl{\"a}schner},
		\citenamefont {Sengstock}, \citenamefont {Goldman},\ and\ \citenamefont
		{Weitenberg}}]{asteria_measuring_2019}%
	\BibitemOpen
	\bibfield  {author} {\bibinfo {author} {\bibfnamefont {L.}~\bibnamefont
			{Asteria}}, \bibinfo {author} {\bibfnamefont {D.~T.}\ \bibnamefont {Tran}},
		\bibinfo {author} {\bibfnamefont {T.}~\bibnamefont {Ozawa}}, \bibinfo
		{author} {\bibfnamefont {M.}~\bibnamefont {Tarnowski}}, \bibinfo {author}
		{\bibfnamefont {B.~S.}\ \bibnamefont {Rem}}, \bibinfo {author} {\bibfnamefont
			{N.}~\bibnamefont {Fl{\"a}schner}}, \bibinfo {author} {\bibfnamefont
			{K.}~\bibnamefont {Sengstock}}, \bibinfo {author} {\bibfnamefont
			{N.}~\bibnamefont {Goldman}}, \ and\ \bibinfo {author} {\bibfnamefont
			{C.}~\bibnamefont {Weitenberg}},\ }\href {\doibase 10.1038/s41567-019-0417-8}
	{\bibfield  {journal} {\bibinfo  {journal} {Nature Physics}\ }\textbf
		{\bibinfo {volume} {15}},\ \bibinfo {pages} {449} (\bibinfo {year}
		{2019})}\BibitemShut {NoStop}%
	\bibitem [{\citenamefont {Roushan}\ \emph {et~al.}(2017)\citenamefont
		{Roushan}, \citenamefont {Neill}, \citenamefont {Megrant}, \citenamefont
		{Chen}, \citenamefont {Babbush}, \citenamefont {Barends}, \citenamefont
		{Campbell}, \citenamefont {Chen}, \citenamefont {Chiaro}, \citenamefont
		{Dunsworth}, \citenamefont {Fowler}, \citenamefont {Jeffrey}, \citenamefont
		{Kelly}, \citenamefont {Lucero}, \citenamefont {Mutus}, \citenamefont
		{O'Malley}, \citenamefont {Neeley}, \citenamefont {Quintana}, \citenamefont
		{Sank}, \citenamefont {Vainsencher}, \citenamefont {Wenner}, \citenamefont
		{White}, \citenamefont {Kapit}, \citenamefont {Neven},\ and\ \citenamefont
		{Martinis}}]{roushan_chiral_2017}%
	\BibitemOpen
	\bibfield  {author} {\bibinfo {author} {\bibfnamefont {P.}~\bibnamefont
			{Roushan}}, \bibinfo {author} {\bibfnamefont {C.}~\bibnamefont {Neill}},
		\bibinfo {author} {\bibfnamefont {A.}~\bibnamefont {Megrant}}, \bibinfo
		{author} {\bibfnamefont {Y.}~\bibnamefont {Chen}}, \bibinfo {author}
		{\bibfnamefont {R.}~\bibnamefont {Babbush}}, \bibinfo {author} {\bibfnamefont
			{R.}~\bibnamefont {Barends}}, \bibinfo {author} {\bibfnamefont
			{B.}~\bibnamefont {Campbell}}, \bibinfo {author} {\bibfnamefont
			{Z.}~\bibnamefont {Chen}}, \bibinfo {author} {\bibfnamefont {B.}~\bibnamefont
			{Chiaro}}, \bibinfo {author} {\bibfnamefont {A.}~\bibnamefont {Dunsworth}},
		\bibinfo {author} {\bibfnamefont {A.}~\bibnamefont {Fowler}}, \bibinfo
		{author} {\bibfnamefont {E.}~\bibnamefont {Jeffrey}}, \bibinfo {author}
		{\bibfnamefont {J.}~\bibnamefont {Kelly}}, \bibinfo {author} {\bibfnamefont
			{E.}~\bibnamefont {Lucero}}, \bibinfo {author} {\bibfnamefont
			{J.}~\bibnamefont {Mutus}}, \bibinfo {author} {\bibfnamefont {P.~J.~J.}\
			\bibnamefont {O'Malley}}, \bibinfo {author} {\bibfnamefont {M.}~\bibnamefont
			{Neeley}}, \bibinfo {author} {\bibfnamefont {C.}~\bibnamefont {Quintana}},
		\bibinfo {author} {\bibfnamefont {D.}~\bibnamefont {Sank}}, \bibinfo {author}
		{\bibfnamefont {A.}~\bibnamefont {Vainsencher}}, \bibinfo {author}
		{\bibfnamefont {J.}~\bibnamefont {Wenner}}, \bibinfo {author} {\bibfnamefont
			{T.}~\bibnamefont {White}}, \bibinfo {author} {\bibfnamefont
			{E.}~\bibnamefont {Kapit}}, \bibinfo {author} {\bibfnamefont
			{H.}~\bibnamefont {Neven}}, \ and\ \bibinfo {author} {\bibfnamefont
			{J.}~\bibnamefont {Martinis}},\ }\href {\doibase 10.1038/nphys3930}
	{\bibfield  {journal} {\bibinfo  {journal} {Nature Physics}\ }\textbf
		{\bibinfo {volume} {13}},\ \bibinfo {pages} {146} (\bibinfo {year}
		{2017})}\BibitemShut {NoStop}%
	\bibitem [{\citenamefont {Mukherjee}\ \emph {et~al.}(2018)\citenamefont
		{Mukherjee}, \citenamefont {Di~Liberto}, \citenamefont {{\"O}hberg},
		\citenamefont {Thomson},\ and\ \citenamefont
		{Goldman}}]{mukherjee_experimental_2018}%
	\BibitemOpen
	\bibfield  {author} {\bibinfo {author} {\bibfnamefont {S.}~\bibnamefont
			{Mukherjee}}, \bibinfo {author} {\bibfnamefont {M.}~\bibnamefont
			{Di~Liberto}}, \bibinfo {author} {\bibfnamefont {P.}~\bibnamefont
			{{\"O}hberg}}, \bibinfo {author} {\bibfnamefont {R.~R.}\ \bibnamefont
			{Thomson}}, \ and\ \bibinfo {author} {\bibfnamefont {N.}~\bibnamefont
			{Goldman}},\ }\href {\doibase 10.1103/PhysRevLett.121.075502} {\bibfield
		{journal} {\bibinfo  {journal} {Physical Review Letters}\ }\textbf {\bibinfo
			{volume} {121}},\ \bibinfo {pages} {075502} (\bibinfo {year}
		{2018})}\BibitemShut {NoStop}%
	\bibitem [{\citenamefont {Mielenz}\ \emph {et~al.}(2016)\citenamefont
		{Mielenz}, \citenamefont {Kalis}, \citenamefont {Wittemer}, \citenamefont
		{Hakelberg}, \citenamefont {Warring}, \citenamefont {Schmied}, \citenamefont
		{Blain}, \citenamefont {Maunz}, \citenamefont {Moehring}, \citenamefont
		{Leibfried},\ and\ \citenamefont {Schaetz}}]{mielenz_arrays_2016}%
	\BibitemOpen
	\bibfield  {author} {\bibinfo {author} {\bibfnamefont {M.}~\bibnamefont
			{Mielenz}}, \bibinfo {author} {\bibfnamefont {H.}~\bibnamefont {Kalis}},
		\bibinfo {author} {\bibfnamefont {M.}~\bibnamefont {Wittemer}}, \bibinfo
		{author} {\bibfnamefont {F.}~\bibnamefont {Hakelberg}}, \bibinfo {author}
		{\bibfnamefont {U.}~\bibnamefont {Warring}}, \bibinfo {author} {\bibfnamefont
			{R.}~\bibnamefont {Schmied}}, \bibinfo {author} {\bibfnamefont
			{M.}~\bibnamefont {Blain}}, \bibinfo {author} {\bibfnamefont
			{P.}~\bibnamefont {Maunz}}, \bibinfo {author} {\bibfnamefont {D.~L.}\
			\bibnamefont {Moehring}}, \bibinfo {author} {\bibfnamefont {D.}~\bibnamefont
			{Leibfried}}, \ and\ \bibinfo {author} {\bibfnamefont {T.}~\bibnamefont
			{Schaetz}},\ }\href {\doibase 10.1038/ncomms11839} {\bibfield  {journal}
		{\bibinfo  {journal} {Nature Communications}\ }\textbf {\bibinfo {volume}
			{7}},\ \bibinfo {pages} {11839} (\bibinfo {year} {2016})}\BibitemShut
	{NoStop}%
	\bibitem [{sup()}]{sup}%
	\BibitemOpen
	\href@noop {} {}\bibinfo {note} {See Supplemental Material for further
		details}\BibitemShut {NoStop}%
	\bibitem [{\citenamefont {Kalis}(2017)}]{kalis_initialization_2017}%
	\BibitemOpen
	\bibfield  {author} {\bibinfo {author} {\bibfnamefont {H.}~\bibnamefont
			{Kalis}},\ }\emph {\bibinfo {title} {Initialization of Quantum States in a
			Two-Dimensional Ion-Trap Array}},\ \href@noop {} {Ph.D. thesis} (\bibinfo
	{year} {2017})\BibitemShut {NoStop}%
	\bibitem [{\citenamefont {Friedenauer}\ \emph {et~al.}(2006)\citenamefont
		{Friedenauer}, \citenamefont {Markert}, \citenamefont {Schmitz},
		\citenamefont {Petersen}, \citenamefont {Kahra}, \citenamefont {Herrmann},
		\citenamefont {Udem}, \citenamefont {H{\"a}nsch},\ and\ \citenamefont
		{Sch{\"a}tz}}]{friedenauer_high_2006}%
	\BibitemOpen
	\bibfield  {author} {\bibinfo {author} {\bibfnamefont {A.}~\bibnamefont
			{Friedenauer}}, \bibinfo {author} {\bibfnamefont {F.}~\bibnamefont
			{Markert}}, \bibinfo {author} {\bibfnamefont {H.}~\bibnamefont {Schmitz}},
		\bibinfo {author} {\bibfnamefont {L.}~\bibnamefont {Petersen}}, \bibinfo
		{author} {\bibfnamefont {S.}~\bibnamefont {Kahra}}, \bibinfo {author}
		{\bibfnamefont {M.}~\bibnamefont {Herrmann}}, \bibinfo {author}
		{\bibfnamefont {T.}~\bibnamefont {Udem}}, \bibinfo {author} {\bibfnamefont
			{T.}~\bibnamefont {H{\"a}nsch}}, \ and\ \bibinfo {author} {\bibfnamefont
			{T.}~\bibnamefont {Sch{\"a}tz}},\ }\href {\doibase 10.1007/s00340-006-2274-2}
	{\bibfield  {journal} {\bibinfo  {journal} {Applied Physics B}\ }\textbf
		{\bibinfo {volume} {84}},\ \bibinfo {pages} {371} (\bibinfo {year}
		{2006})}\BibitemShut {NoStop}%
	\bibitem [{\citenamefont {Grifoni}\ and\ \citenamefont
		{Haenggi}(1998)}]{grifoni_driven_1998}%
	\BibitemOpen
	\bibfield  {author} {\bibinfo {author} {\bibfnamefont {M.}~\bibnamefont
			{Grifoni}}\ and\ \bibinfo {author} {\bibfnamefont {P.}~\bibnamefont
			{Haenggi}},\ }\href {\doibase 10.1016/S0370-1573(98)00022-2} {\bibfield
		{journal} {\bibinfo  {journal} {Physics Reports}\ }\textbf {\bibinfo {volume}
			{304}},\ \bibinfo {pages} {229} (\bibinfo {year} {1998})}\BibitemShut
	{NoStop}%
	\bibitem [{Note1()}]{Note1}%
	\BibitemOpen
	\bibinfo {note} {Note: the phase relation of the ion oscillators results from
		different constant phase offsets, e.g. caused by supply wiring of the control
		electrodes, anharmonic contributions of the trapping potential or the
		duration between the start of excitation and modulation
		potentials.}\BibitemShut {Stop}%
	\bibitem [{\citenamefont {Hite}\ \emph {et~al.}(2012)\citenamefont {Hite},
		\citenamefont {Colombe}, \citenamefont {Wilson}, \citenamefont {Brown},
		\citenamefont {Warring}, \citenamefont {J{\"o}rdens}, \citenamefont {Jost},
		\citenamefont {McKay}, \citenamefont {Pappas}, \citenamefont {Leibfried},\
		and\ \citenamefont {Wineland}}]{hite_100-fold_2012}%
	\BibitemOpen
	\bibfield  {author} {\bibinfo {author} {\bibfnamefont {D.~A.}\ \bibnamefont
			{Hite}}, \bibinfo {author} {\bibfnamefont {Y.}~\bibnamefont {Colombe}},
		\bibinfo {author} {\bibfnamefont {A.~C.}\ \bibnamefont {Wilson}}, \bibinfo
		{author} {\bibfnamefont {K.~R.}\ \bibnamefont {Brown}}, \bibinfo {author}
		{\bibfnamefont {U.}~\bibnamefont {Warring}}, \bibinfo {author} {\bibfnamefont
			{R.}~\bibnamefont {J{\"o}rdens}}, \bibinfo {author} {\bibfnamefont {J.~D.}\
			\bibnamefont {Jost}}, \bibinfo {author} {\bibfnamefont {K.~S.}\ \bibnamefont
			{McKay}}, \bibinfo {author} {\bibfnamefont {D.~P.}\ \bibnamefont {Pappas}},
		\bibinfo {author} {\bibfnamefont {D.}~\bibnamefont {Leibfried}}, \ and\
		\bibinfo {author} {\bibfnamefont {D.~J.}\ \bibnamefont {Wineland}},\ }\href
	{\doibase 10.1103/PhysRevLett.109.103001} {\bibfield  {journal} {\bibinfo
			{journal} {Physical Review Letters}\ }\textbf {\bibinfo {volume} {109}},\
		\bibinfo {pages} {103001} (\bibinfo {year} {2012})}\BibitemShut {NoStop}%
	\bibitem [{\citenamefont {Labaziewicz}\ \emph {et~al.}(2008)\citenamefont
		{Labaziewicz}, \citenamefont {Ge}, \citenamefont {Antohi}, \citenamefont
		{Leibrandt}, \citenamefont {Brown},\ and\ \citenamefont
		{Chuang}}]{labaziewicz_suppression_2008}%
	\BibitemOpen
	\bibfield  {author} {\bibinfo {author} {\bibfnamefont {J.}~\bibnamefont
			{Labaziewicz}}, \bibinfo {author} {\bibfnamefont {Y.}~\bibnamefont {Ge}},
		\bibinfo {author} {\bibfnamefont {P.}~\bibnamefont {Antohi}}, \bibinfo
		{author} {\bibfnamefont {D.}~\bibnamefont {Leibrandt}}, \bibinfo {author}
		{\bibfnamefont {K.~R.}\ \bibnamefont {Brown}}, \ and\ \bibinfo {author}
		{\bibfnamefont {I.~L.}\ \bibnamefont {Chuang}},\ }\href {\doibase
		10.1103/PhysRevLett.100.013001} {\bibfield  {journal} {\bibinfo  {journal}
			{Physical Review Letters}\ }\textbf {\bibinfo {volume} {100}},\ \bibinfo
		{pages} {013001} (\bibinfo {year} {2008})}\BibitemShut {NoStop}%
	\bibitem [{\citenamefont {Schaetz}\ \emph {et~al.}(2007)\citenamefont
		{Schaetz}, \citenamefont {Friedenauer}, \citenamefont {Schmitz},
		\citenamefont {Petersen},\ and\ \citenamefont
		{Kahra}}]{schaetz_towards_2007}%
	\BibitemOpen
	\bibfield  {author} {\bibinfo {author} {\bibfnamefont {T.}~\bibnamefont
			{Schaetz}}, \bibinfo {author} {\bibfnamefont {A.}~\bibnamefont
			{Friedenauer}}, \bibinfo {author} {\bibfnamefont {H.}~\bibnamefont
			{Schmitz}}, \bibinfo {author} {\bibfnamefont {L.}~\bibnamefont {Petersen}}, \
		and\ \bibinfo {author} {\bibfnamefont {S.}~\bibnamefont {Kahra}},\ }\href
	{\doibase 10.1080/09500340701639631} {\bibfield  {journal} {\bibinfo
			{journal} {Journal of Modern Optics}\ }\textbf {\bibinfo {volume} {54}},\
		\bibinfo {pages} {2317} (\bibinfo {year} {2007})}\BibitemShut {NoStop}%
	\bibitem [{\citenamefont {Schaetz}\ \emph {et~al.}(2013)\citenamefont
		{Schaetz}, \citenamefont {Monroe},\ and\ \citenamefont
		{Esslinger}}]{schaetz_focus_2013}%
	\BibitemOpen
	\bibfield  {author} {\bibinfo {author} {\bibfnamefont {T.}~\bibnamefont
			{Schaetz}}, \bibinfo {author} {\bibfnamefont {C.~R.}\ \bibnamefont {Monroe}},
		\ and\ \bibinfo {author} {\bibfnamefont {T.}~\bibnamefont {Esslinger}},\
	}\href {\doibase 10.1088/1367-2630/15/8/085009} {\bibfield  {journal}
		{\bibinfo  {journal} {New Journal of Physics}\ }\textbf {\bibinfo {volume}
			{15}},\ \bibinfo {pages} {085009} (\bibinfo {year} {2013})}\BibitemShut
	{NoStop}%
	\bibitem [{\citenamefont {Huber}\ and\ \citenamefont
		{Lindner}(2011)}]{huber_topological_2011}%
	\BibitemOpen
	\bibfield  {author} {\bibinfo {author} {\bibfnamefont {S.~D.}\ \bibnamefont
			{Huber}}\ and\ \bibinfo {author} {\bibfnamefont {N.~H.}\ \bibnamefont
			{Lindner}},\ }\href {\doibase 10.1073/pnas.1110813108} {\bibfield  {journal}
		{\bibinfo  {journal} {Proceedings of the National Academy of Sciences}\
		}\textbf {\bibinfo {volume} {108}},\ \bibinfo {pages} {19925} (\bibinfo
		{year} {2011})}\BibitemShut {NoStop}%
	\bibitem [{\citenamefont {Lemmer}\ \emph {et~al.}(2018)\citenamefont {Lemmer},
		\citenamefont {Cormick}, \citenamefont {Tamascelli}, \citenamefont {Schaetz},
		\citenamefont {Huelga},\ and\ \citenamefont
		{Plenio}}]{lemmer_trapped-ion_2018}%
	\BibitemOpen
	\bibfield  {author} {\bibinfo {author} {\bibfnamefont {A.}~\bibnamefont
			{Lemmer}}, \bibinfo {author} {\bibfnamefont {C.}~\bibnamefont {Cormick}},
		\bibinfo {author} {\bibfnamefont {D.}~\bibnamefont {Tamascelli}}, \bibinfo
		{author} {\bibfnamefont {T.}~\bibnamefont {Schaetz}}, \bibinfo {author}
		{\bibfnamefont {S.~F.}\ \bibnamefont {Huelga}}, \ and\ \bibinfo {author}
		{\bibfnamefont {M.~B.}\ \bibnamefont {Plenio}},\ }\href {\doibase
		10.1088/1367-2630/aac87d} {\bibfield  {journal} {\bibinfo  {journal} {New
				Journal of Physics}\ }\textbf {\bibinfo {volume} {20}},\ \bibinfo {pages}
		{073002} (\bibinfo {year} {2018})}\BibitemShut {NoStop}%
	\bibitem [{\citenamefont {Bermudez}\ \emph {et~al.}(2013)\citenamefont
		{Bermudez}, \citenamefont {Bruderer},\ and\ \citenamefont
		{Plenio}}]{bermudez_controlling_2013}%
	\BibitemOpen
	\bibfield  {author} {\bibinfo {author} {\bibfnamefont {A.}~\bibnamefont
			{Bermudez}}, \bibinfo {author} {\bibfnamefont {M.}~\bibnamefont {Bruderer}},
		\ and\ \bibinfo {author} {\bibfnamefont {M.~B.}\ \bibnamefont {Plenio}},\
	}\href {\doibase 10.1103/PhysRevLett.111.040601} {\bibfield  {journal}
		{\bibinfo  {journal} {Physical Review Letters}\ }\textbf {\bibinfo {volume}
			{111}},\ \bibinfo {pages} {040601} (\bibinfo {year} {2013})}\BibitemShut
	{NoStop}%
	\bibitem [{\citenamefont {Peano}\ \emph {et~al.}(2016)\citenamefont {Peano},
		\citenamefont {Houde}, \citenamefont {Marquardt},\ and\ \citenamefont
		{Clerk}}]{peano_topological_2016}%
	\BibitemOpen
	\bibfield  {author} {\bibinfo {author} {\bibfnamefont {V.}~\bibnamefont
			{Peano}}, \bibinfo {author} {\bibfnamefont {M.}~\bibnamefont {Houde}},
		\bibinfo {author} {\bibfnamefont {F.}~\bibnamefont {Marquardt}}, \ and\
		\bibinfo {author} {\bibfnamefont {A.~A.}\ \bibnamefont {Clerk}},\ }\href
	{\doibase 10.1103/PhysRevX.6.041026} {\bibfield  {journal} {\bibinfo
			{journal} {Physical Review X}\ }\textbf {\bibinfo {volume} {6}},\ \bibinfo
		{pages} {041026} (\bibinfo {year} {2016})}\BibitemShut {NoStop}%
	\bibitem [{\citenamefont {Ozawa}\ \emph {et~al.}(2019)\citenamefont {Ozawa},
		\citenamefont {Price}, \citenamefont {Amo}, \citenamefont {Goldman},
		\citenamefont {Hafezi}, \citenamefont {Lu}, \citenamefont {Rechtsman},
		\citenamefont {Schuster}, \citenamefont {Simon}, \citenamefont {Zilberberg},\
		and\ \citenamefont {Carusotto}}]{ozawa_topological_2019}%
	\BibitemOpen
	\bibfield  {author} {\bibinfo {author} {\bibfnamefont {T.}~\bibnamefont
			{Ozawa}}, \bibinfo {author} {\bibfnamefont {H.~M.}\ \bibnamefont {Price}},
		\bibinfo {author} {\bibfnamefont {A.}~\bibnamefont {Amo}}, \bibinfo {author}
		{\bibfnamefont {N.}~\bibnamefont {Goldman}}, \bibinfo {author} {\bibfnamefont
			{M.}~\bibnamefont {Hafezi}}, \bibinfo {author} {\bibfnamefont
			{L.}~\bibnamefont {Lu}}, \bibinfo {author} {\bibfnamefont {M.~C.}\
			\bibnamefont {Rechtsman}}, \bibinfo {author} {\bibfnamefont {D.}~\bibnamefont
			{Schuster}}, \bibinfo {author} {\bibfnamefont {J.}~\bibnamefont {Simon}},
		\bibinfo {author} {\bibfnamefont {O.}~\bibnamefont {Zilberberg}}, \ and\
		\bibinfo {author} {\bibfnamefont {I.}~\bibnamefont {Carusotto}},\ }\href
	{\doibase 10.1103/RevModPhys.91.015006} {\bibfield  {journal} {\bibinfo
			{journal} {Reviews of Modern Physics}\ }\textbf {\bibinfo {volume} {91}},\
		\bibinfo {pages} {015006} (\bibinfo {year} {2019})}\BibitemShut {NoStop}%
	\bibitem [{\citenamefont {Porras}\ and\ \citenamefont
		{{Fern{\'a}ndez-Lorenzo}}(2019)}]{porras_topological_2019}%
	\BibitemOpen
	\bibfield  {author} {\bibinfo {author} {\bibfnamefont {D.}~\bibnamefont
			{Porras}}\ and\ \bibinfo {author} {\bibfnamefont {S.}~\bibnamefont
			{{Fern{\'a}ndez-Lorenzo}}},\ }\href {\doibase 10.1103/PhysRevLett.122.143901}
	{\bibfield  {journal} {\bibinfo  {journal} {Physical Review Letters}\
		}\textbf {\bibinfo {volume} {122}},\ \bibinfo {pages} {143901} (\bibinfo
		{year} {2019})}\BibitemShut {NoStop}%
	\bibitem [{\citenamefont {Bermudez}\ and\ \citenamefont
		{Porras}(2015)}]{bermudez_interaction-dependent_2015}%
	\BibitemOpen
	\bibfield  {author} {\bibinfo {author} {\bibfnamefont {A.}~\bibnamefont
			{Bermudez}}\ and\ \bibinfo {author} {\bibfnamefont {D.}~\bibnamefont
			{Porras}},\ }\href {\doibase 10.1088/1367-2630/17/10/103021} {\bibfield
		{journal} {\bibinfo  {journal} {New Journal of Physics}\ }\textbf {\bibinfo
			{volume} {17}},\ \bibinfo {pages} {103021} (\bibinfo {year}
		{2015})}\BibitemShut {NoStop}%
\end{thebibliography}
%

%

\appendix
\clearpage


\section*{Supplementary material}

# Supplementary material: Floquet-engineered vibrational dynamics in a two-dimensional array of trapped ions

Philip Kiefer,[1,*] Frederick Hakelberg,[1] Matthias Wittemer,[1]
Alejandro Bermúdez,[2,†] Diego Porras,[3,‡] Ulrich Warring,[1] and Tobias Schaetz[1]

[1]*Albert-Ludwigs-Universität Freiburg, Physikalisches Institut,*
*Hermann-Herder-Strasse 3, 79104 Freiburg, Germany*
[2]*Departamento de Física Teórica, Universidad Complutense, 28040 Madrid, Spain*
[3]*Instituto de Física Fundamental IFF-CSIC,*
*Calle Serrano 113b, 28006 Madrid, Spain*

## VIBRATIONAL DYNAMICS HAMILTONIAN

As proposed in [6, 7], a parametric modulation of the individual motional frequencies can induce effective vibrational dynamics described by the Hamiltonian

$$H = \sum_{\mathbf{j}>\mathbf{k}} \left( J_{\mathbf{jk}}^{\text{eff}} e^{i\Phi_{\mathbf{jk}}^{\text{P}}} a_{\mathbf{j}}^\dagger a_{\mathbf{k}} + \text{H.c.} \right) + \sum_{\mathbf{j}} V_{\mathbf{j}}(a_{\mathbf{j}}^\dagger, a_{\mathbf{j}}). \qquad (1)$$

We have introduced bosonic operators that create-annihilate vibrational quanta (phonons) $a_{\mathbf{j}}^\dagger, a_{\mathbf{j}}$ at $T_{\mathbf{j}}$ of the 2D array. Hamiltonian (1) describes the vibrational dynamics in terms of the hopping of localized phonons between different ions. $J_{\mathbf{jk}}^{\text{eff}}$ defines the Floquet-engineered tunneling strength, and $\Phi_{\mathbf{jk}}^{\text{P}}$ is the so-called Peierls phase, which can mimic the coupling of a charged particle to an external electromagnetic field. We note that by modifying the strength of the parametric drive, we can tune $J_{\mathbf{jk}}^{\text{eff}}$, while $\Phi_{\mathbf{jk}}^{\text{P}}$ solely depends on the phases $\varphi_{\mathbf{j}}$ of the parametric drive of each individual trap. The exchange rate, which is directly accessible in the experiment, corresponds to $\Omega_{\text{AC}} = 2 \cdot J_{\mathbf{jk}}^{\text{eff}}$. Finally, $V_{\mathbf{j}}(a_{\mathbf{j}}^\dagger, a_{\mathbf{j}})$ includes single-site terms, such as additional anharmonicities that can be interpreted as local phonon-phonon interactions.

## PLANAR SURFACE-ELECTRODE TRAP ARRAY

We use a planar surface-electrode ion-trap array [1] with an electrode geometry optimized for isolation from the environment, inter-site coupling, and individual site control [2]. The trapping potential $\phi_{\text{trap}}$ is based on a global radio-frequency (rf) potential $\phi_{\text{RF}}$ and further tuned by local control potentials $\phi^{\text{C}}$. Two rf electrodes are supplied in-phase with a zero-to-peak amplitude $U_{\text{RF}} \simeq 40\,\text{V}$, oscillating at $\Omega_{\text{RF}}/(2\pi) \simeq 89\,\text{MHz}$. Distinct control potentials $\phi^{\text{C}}$ are generated by 30 control electrodes, each connected to an arbitrary waveform generator with an output voltage of $\pm 10\,\text{V}$ and an update rate of $2\pi \times 50\,\text{MHz}$ [3]. Initially, we evaporate Mg atoms below loading holes in the trap chip and photoionize near the $T_{\mathbf{j}}$. To fill the required trapping sites, we relocate ions between the $T_{\mathbf{j}}$ using dedicated control potentials. Each site can be loaded with single or multiple ions and we tune typical center of mass eigenfrequencies within the range of $2\pi \times (2-10)\,\text{MHz}$.

**EXPERIMENTAL REALIZATION OF THE PERIODIC MODULATION**

In general, we could modulate the motional frequencies by periodically oscillating control curvatures [1]. Here, we realize the periodic modulation by exploitation of anharmonic third-order contributions of the trapping potential. Due to these anharmonic contributions, the motional eigenfrequencies depend on the position of the ion. By application of a control potential $\phi^{\mathcal{M}}$ with amplitude $u^{\mathcal{M}}$ the ion becomes displaced (up to $\simeq 1\,\mu\text{m}$) and therefore its eigenfrequency changes. We characterize static effects of $\phi^{\mathcal{M}}$ on the lowest eigenfrequency $\omega_{\mathbf{j}}$ by

$$\omega_{\mathbf{j}}(u^{\mathcal{M}}) = c_{\mathbf{j}} + b_{\mathbf{j}} \cdot u^{\mathcal{M}} + a_{\mathbf{j}} \cdot (u^{\mathcal{M}})^2. \tag{2}$$

The harmonic contribution of the trapping potential is described by $c_{\mathbf{j}}$, higher third and fourth order contributions are characterized by $b_{\mathbf{j}}$ and $a_{\mathbf{j}}$, respectively.

Exemplary, we show measurement results for single ions in $T_0$ (blue) and $T_1$ (orange), and corresponding model fits in Fig. S1. To realize a near sinusoidal periodic modulation of the motional eigenfrequency, we require sufficient third order contributions $b_{\mathbf{j}}$. An oscillating control potential $\phi^{\mathcal{M}} \propto u^{\mathcal{M}} \cdot \sin(\Omega^{\mathcal{M}} t)$ leads in this case to a frequency modulation, with modulation index $\eta_{\mathbf{j}} = u^{\mathcal{M}} \cdot |b_{\mathbf{j}}|/\Omega^{\mathcal{M}}$ and modulation frequency $\Omega^{\mathcal{M}}$. Fourth order contributions of the local potentials $a_{\mathbf{j}}$ lead to an observable shift of the average eigenfrequency by $\delta_{\text{car},\mathbf{j}} \propto a_{\mathbf{j}}(u^{\mathcal{M}})^2$, while we can neglect effects on the modulation index $\eta_{\mathbf{j}}$. During all performed experimental sequences, we calibrate frequency shifts of $\omega_{\mathbf{j}}$ due to $\phi_{\mathbf{j}}^{\mathcal{M}}$, up to $|\delta_{\text{car},\mathbf{j}}|/(2\pi) \simeq 30\,\text{kHz}$. In general we can also modulate with a waveform that countervails the fourth order effects.

The modulation crosstalk between trapping sites depend on the static local potential at each site and the chosen modulation potential $\phi^{\mathcal{M}}$. To minimize crosstalk between the three sites, we modulate with $\phi_{\mathbf{j}}^{\mathcal{M}}$ applied up to three electrodes.

**Model to derive modulation indices**

In absence of the modulation, the frequency spectrum of the coherent motional excitation of an ion is described by the Fourier transform of the excitation pulse. Here the rectangular shaped excitation pulse leads to a *sinc*-shaped response of the ion [5]. The presence of the modulation causes characteristic sidebands. To determine the modulation indices $\eta_{\mathbf{j}}$ (see

3

Fig. 1) we model the motional excitation E of the ion with the function:

$$E(\eta_{\mathbf{j}}) = \sum_{m=-\infty}^{\infty} \tilde{E}_{\mathbf{j}} \cdot \mathcal{J}_m(\eta_{\mathbf{j}})^2 \cdot \text{sinc}(t^{\mathcal{M}}(\Omega^{\mathcal{E}} - (\widetilde{\omega}_{\mathbf{j}} + m \cdot \Omega^{\mathcal{M}} + \delta_{\text{car},\mathbf{j}}))/2)^2 \quad (3)$$

where $\mathcal{J}_m(\eta_{\mathbf{j}})$ are Bessel functions of the first kind. The amplitude of the unmodulated coherent state is represented by $\tilde{E}_{\mathbf{j}}$, the ion's unmodulated eigenfrequency by $\widetilde{\omega}_{\mathbf{j}}$ and the shift of the center frequency by $\delta_{\text{car},\mathbf{j}}$. The modulation frequency is given by $\Omega^{\mathcal{M}}$, the excitation frequency by $\Omega^{\mathcal{E}}$. The simultaneous modulation and excitation is applied for a duration $t^{\mathcal{E}} = t^{\mathcal{M}}$.

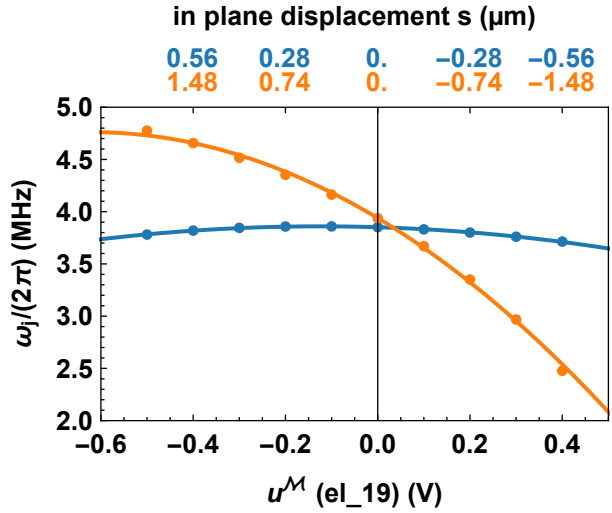

FIG. S1. **Characterization of anharmonic contributions of the trapping potential.** Static measured $\omega_{\mathbf{j}}$ ($\omega_0$ blue, $\omega_1$ orange data points; errorbars are smaller than marker size) depending on the amplitude $u^{\mathcal{M}}$ of $\phi^{\mathcal{M}}$, generated by a single electrode ('el_19', see [4]). The solid lines represent model fits of equation (1), yielding $a_0/(2\pi) = -0.543(9)\,\text{MHz/V}^2$, $b_0/(2\pi) = -0.135(3)\,\text{MHz/V}$, $c_0/(2\pi) = 3.851(1)\,\text{MHz}$, and $a_1/(2\pi) = -2.1(2)\,\text{MHz/V}^2$, $b_1/(2\pi) = -2.65(4)\,\text{MHz/V}$, $c_1/(2\pi) = 3.94(2)\,\text{MHz/V}^2$. The crosstalk of the linear contribution at $u^{\mathcal{M}} = 0\,\text{V}$ is $b_0/b_1 = 5.1(1)\,\%$. The corresponding in-plane displacement at $T_{\mathbf{j}}$ is captured by a camera in a separate measurement. The top x-axis represent the ions' linear modeled displacements s parallel to the surface plane. Static out of plane mode orientations are typically below $25°$, corresponding to a 10% enlarged displacement along its mode axis. We can rewrite the effect of the anharmonicities depending on the displacement s by $\omega_{\mathbf{j}}(s)/(2\pi) = C_{\mathbf{j}} + B_{\mathbf{j}} \cdot s + A_{\mathbf{j}} \cdot s^2$ with $A_0 = -0.280(7)\,\text{MHz/(μm)}^2$, $B_0 = 0.097(2)\,\text{MHz/(μm)}$, $A_1 = -0.15(2)\,\text{MHz/(μm)}^2$, $B_1 = 0.71(4)\,\text{MHz/(μm)}$ and $C_{\mathbf{j}} = c_{\mathbf{j}}$.

4

## MODEL OF COHERENT PHONON EXCHANGE

We model the motional excitation $\bar{n}_\mathbf{j}$ of an initially Doppler cooled ion (with initial $\bar{n}_\mathbf{j} < 20$) at $T_\mathbf{j}$, which is coherently exchanging phonons with other sites by

$$\bar{n}_\mathbf{j}(t_\mathbf{j}^{\mathcal{M}}) = N/2 \cdot (1 - \cos(\Omega_{\text{AC}} \cdot t_\mathbf{j}^{\mathcal{M}}) \cdot e^{-(t_\mathbf{j}^{\mathcal{M}}/\tau)^2}). \qquad (4)$$

The exchange amplitude is given by N, and $t_\mathbf{j}^{\mathcal{M}}$ describes the exchange duration. Instabilities of the eigenfrequencies, on timescales of ensemble averaging, are taken into account by a Gaussian dephasing term $e^{-(t_\mathbf{j}^{\mathcal{M}}/\tau)^2}$ with dephasing timescale $\tau$. The initial contributions $\bar{n}_\mathbf{j}(t_\mathbf{j}^{\mathcal{M}} = 0)$ are not considered for the model fits.

---

* philip.kiefer@physik.uni-freiburg.de; https://www.qsim.uni-freiburg.de

† albermud@fucm.es

‡ d.porras@iff.csic.es

[1] M. Mielenz, H. Kalis, M. Wittemer, F. Hakelberg, U. Warring, R. Schmied, M. Blain, P. Maunz, D. L. Moehring, D. Leibfried, and T. Schaetz, Nature Communications **7**, 11839 (2016).

[2] R. Schmied, J. H. Wesenberg, and D. Leibfried, Physical Review Letters **102**, 233002 (2009).

[3] R. Bowler, U. Warring, J. W. Britton, B. C. Sawyer, and J. Amini, Review of Scientific Instruments **84**, 033108 (2013).

[4] H. Kalis, F. Hakelberg, M. Wittemer, M. Mielenz, U. Warring, and T. Schaetz, Physical Review A **94**, 023401 (2016).

[5] H. Kalis, *Initialization of Quantum States in a Two-Dimensional Ion-Trap Array*, Ph.D. thesis (2017).

[6] A. Bermudez, T. Schaetz, and D. Porras, Physical Review Letters **107**, 150501 (2011).

[7] A. Bermudez, T. Schaetz, and D. Porras, New Journal of Physics **14**, 053049 (2012).



\end{document}